\documentclass[11pt]{article}

\usepackage[a4paper]{geometry}

\usepackage[latin1]{inputenc}
\usepackage{latexsym}
\usepackage{amsmath}
\usepackage{amsthm}
\usepackage{amssymb}

\newtheorem{theorem}{Theorem}
\newtheorem{lemma}{Lemma}

\theoremstyle{remark}
\newtheorem*{remark}{Remark}

\title{On the cylindrically symmetric Einstein-Vlasov system}

\author{Mikael Fjällborg}

\begin{document}

\maketitle
\textbf{Abstract}

For the cylindrically symmetric ''asymptotically flat'' Einstein equations
in the case of electro-vacuum it is known that solutions exist globally and
also that this class of spacetimes is causally geodesically complete. Hence
strong cosmic censorship holds for this class. An interesting question
is whether these results can be generalized to include spacetimes with phenomenological matter, e.g. collisionless matter described by the Vlasov equation. Spherically symmetric asymptotically flat solutions of the Einstein-Vlasov system with small initial data are known to be causally geodesically complete. For arbitrary (in size) data it has been shown that if a singularity occurs, the first one occurs at the center of symmetry. In this paper we begin to study the question of global existence for the cylindrically symmetric Einstein-Vlasov system with general (in size) data and we show that if a singularity occurs at all, the first one occurs at the axis of symmetry.

\bigskip

\section{Introduction}
Consider a large ensemble of particles, e.g. stars in a galaxy, which interact by the gravitational field which they create collectively. In particular, assume that collisions are sufficiently rare to be neglected. In the Newtonian case one arrives at the so called Vlasov-Poisson system. From a mathematical point of view this system is quite well understood. The question of global existence for the Vlasov-Poisson system in three space dimensions for general smooth initial data was first solved by Pfaffelmoser \cite{Pf}. The relativistic analogue is much more complicated due to the fact that the field is governed by the Einstein equations instead of the Poisson equation. This is the so called Einstein-Vlasov system. For a detailed review concerning global results for the Einstein-Vlasov system, see \cite{And2}. 

As is well known, freely falling particles follow the geodesics of spacetime. By assuming that collisions are sufficiently rare to be neglected, the matter distribution function $f$, i.e. a function $f$ defined on phase space which describes the density of particles at position $x$ with momentum $p$ at time $t$, is conserved along the geodesics. Hence $f$ satisfies the Vlasov equation

\begin{equation}
\frac{\partial f}{\partial t}+\frac{p^{\nu}}{p^{0}}\frac{\partial f}{\partial x^{\nu}}-\Gamma_{bc}^{\nu}p^{b}p^{c}\frac{\partial f}{\partial p^{\nu}}=0,\text{ }b,c=0,1,2,3,\text{ }\nu=1,2,3,
\end{equation}

where $\Gamma_{bc}^{\nu}$ are the Christoffel symbols associated to the metric in the local coordinates and $p^{0}$ is a solution to $g_{ab}p^{a}p^{b}=-1$. For more details concerning relativistic kinetic  theory, see e.g. \cite{Ehlers} or \cite{And2}. Now the Vlasov equation is coupled to the Einstein equations 
\begin{equation}
R_{ab}-\frac{1}{2}Rg_{ab}=8\pi T_{ab},
\end{equation}
where the energy-momentum tensor $T_{ab}$ is given by

\begin{equation}
T_{ab}=-\int_{\mathbb{R}^{3}}f(t,x,p)\frac{p_{a}p_{b}}{p_{0}}\left |g\right |^{\frac{1}{2}}dp.
\end{equation}

Here $t=x^{0}$, $x=(x^{1},x^{2},x^{3})$, $p=(p^{1},p^{2},p^{3})$, $\left |g\right |=\det\left (g_{ab}\right ),\text{ }a,b=0,1,2,3$ and $c=G=1$ where $c$ is the velocity of light and $G$ is the gravitational constant.

When an isolated body is studied, the spacetime is assumed to be flat ``far away'' from the body. These spacetimes are called asymptotically flat. The topology of the initial surface is $\mathbb{R}^{3}$ and the metric has to satisfy certain fall-off conditions. Intuitively the metric should tend to the Minkowski metric far away from the body in some coordinate system. This class is compatible only with a few different symmetry classes where spherical- and axial symmetry are the most important ones. In the class of asymptotically flat spacetimes only the spherically symmetric case has been considered. In \cite{RR1}, \cite{RR2} a global existence result for the spherically symmetric case with small initial data is obtained and it is proved that these spacetimes are geodesically complete. In \cite{RRS} it is shown, for arbitrary (in size) initial data, that if a singularity occurs, the first one occurs at the center of symmetry. The purpose of this paper is to prove a similar result for cylindrically symmetric spacetimes.

Cylindrically symmetric spacetimes, admit asymptotic flatness conditions in two directions of the initial surface and is therefore not asymptotically flat. The reason that cylindrical spacetimes do not belong to the class of asymptotically flat spacetimes is that the symmetry group contains translation symmetry. However following Berger, Chrusciel and Moncrief \cite{BCM} we can state an asymptotic flatness condition in two of the three space dimensions on the initial surface (see section 2 below for a definition). For cylindrically symmetric, asymptotically flat electro-vacuum spacetimes it is known that global existence holds \cite{BCM}, and in addition that these spacetimes are causally geodesically complete. This is done by rewriting the Einstein equations as a rotationally symmetric wavemap from $\mathbb{R}^{1+2}$ into a Riemannian manifold. The target manifold satisfies a certain condition on the behaviour of the curvature at infinity and a convexity condition. Hence the results of Christodoulou and Tahvildar-Zadeh \cite{CT1,CT2} apply. When matter is introduced this wavemap structure is lost. However it is possible to rewrite the Einstein equations as a wavemap with a forcing term from a rotationally symmetric $2+1$-dimensional curved spacetime with the same target manifold as in the vacuum case. The forcing term corresponds to the matter term. Unfortunately we are not able to use this result and generalize the Christodoulou/Tahvildar-Zadeh result to this type of wavemap. Instead we use the method of Andreasson \cite{And1}, for cosmological spacetimes. In \cite{And1} there is no center of symmetry and by using that method we are not able to control the field or the matter at the axis of symmetry. Hence we need to introduce conditions both on the field and on the matter distribution function in a vicinity of the axis of symmetry to obtain global existence. However in the so called polarized case, i.e. when the metric is diagonal, we are able to remove one of the assumptions and still obtain the same result by keeping the condition on the matter. This is in analogy with the spherically symmetric case in \cite{RRS}. The outline of the paper is as follows. In section 2 we define cylindrical spacetimes and the corresponding definition of asymptotic flatness for this case. In section 3 the main theorems are stated and the proofs of the theorems are outlined in section 4 and 5 respectively. Theorem 2 is proved by the strategy outlined in \cite{And1}. However to bound the matter terms we need to introduce a rescaled momentum. Theorem 1 is proved by Sobolev methods combined with the proof of Theorem 2.

\section{Cylindrical symmetry and asymptotic flatness}

By cylindrically symmetric initial data we mean that there exists a
2-dimensional isometry group $G\cong U(1)\times \mathbb{R},$ acting on a spacelike
initial surface $S\cong \mathbb{R}^{3}$. The metric $h_{\mu \nu }$ and the
second fundamental form $K_{\mu \nu }$ of $S$ satisfy%
\begin{equation*}
\pounds _{X^{a}}h_{\mu \nu }=\pounds _{X^{a}}K_{\mu \nu }=0,
\end{equation*}%
where $X^{a},$ $a=2,3$ are two Killing vectors that generate the isometry
group. Furthermore the matter distribution function $f_{0}$ is also
invariant under the action of $G,$ i.e. $f_{0}(\Omega x,\Omega v)=$ $%
f_{0}(x,v),$ $\forall x,v\in \mathbb{R}^{3}$ where $\Omega \in G.$ 

The notion of asymptotic flatness is subtle. Intuitively one can think of an
asymptotically flat metric $g_{ab}$ as a metric that far out tends to the
Minkowski metric in some coordinate system. However this turns out to be a
difficult definition to work with, because the coordinate invariance must be
checked in all statements. It is not clear how to specify in which way the
limits $"r\rightarrow \infty "$ are to be taken in a coordinate independent
manner. We will not give the precise definition of asymptotic flatness here,
but just mention the fact that there are only a few symmetry classes compatible with asymptotic flatness where spherical symmetry and axial symmetry are the most interesting ones. The reason that cylindrically symmetric spacetimes are not compatible with asymptotic flatness is that the action group contains translation symmetry as mentioned in the introduction. Nevertheless, we can use another type of ''asymptotic flatness'' condition because of the absence of translation symmetry in two directions of the 3-dimensional spacelike surface. Following \cite{BCM}, we define asymptotically flat cylindrically symmetric initial data by:

\bigskip

(i) $(S,h_{\mu \nu })$ is geodesically complete with $S$ and $S/G$ not
compact.\bigskip

(ii) There exists a $G$-invariant subset $K$ of $S$ such that $h_{\mu
\nu }$ is flat on $S\backslash K.$\bigskip

(iii) $K/G$ is compact.\bigskip

(iv) $f_{0}\equiv 0,$ in the complement of $K.$

\bigskip

In \cite{BCM} it is proved that the action of an abelian symmetry group
together with this notion of asymptotic flatness and a reasonable energy
condition, which Vlasov matter satisfies, exclude all possible spacelike
surfaces and group actions except for $S\cong \mathbb{R}^{3}$ with cylindrical
symmetry or a periodic identification thereof along the translation axis.

In this paper we are going to use coordinates $t\in \lbrack 0,\infty ),$ $%
r\in \lbrack 0,\infty ),$ $z\in (-\infty ,\infty )$ and $\theta \in \lbrack
0,2\pi ]$ which are adapted to the symmetry such that the two Killing vector fields are%
\begin{equation*}
X^{2}=\frac{\partial }{\partial z},\text{ }X^{3}=\frac{\partial }{\partial
\theta }.
\end{equation*}%
Hence the translation symmetry is along the z-axis and the rotational
symmetry is around the z-axis. We consider a metric of the following form%
\begin{equation}
ds^{2}=-\alpha e^{2\left( \eta -\gamma \right) }dt^{2}+e^{2\left( \eta
-\gamma \right) }dr^{2}+e^{2\gamma }\left( dz+Ad\theta \right)
^{2}+r^{2}e^{-2\gamma }d\theta ^{2},  \label{met}
\end{equation}%
where $\alpha ,$ $\eta ,$ $\gamma $ and $A$ depend on $t$ and $r$. It should be remarked that this is not the most general asymptotically flat, cylindrically symmetric metric possible. However in the elctro-vacuum case it can be shown that a metric of this type is the most general. With this parametrization of the metric the Einstein-Vlasov system reads

\underline{The Einstein-matter constraint equations:}%
\begin{align}
\frac{\eta _{t}}{r}& =2\gamma _{r}\gamma _{t}+\frac{A_{r}A_{t}e^{4\gamma }}{%
2r^{2}}-\sqrt{\alpha }e^{2\left( \eta -\gamma \right) }J,  \label{c1} \\
\frac{\eta _{r}}{r}& =\gamma _{r}^{2}+\frac{\gamma _{t}^{2}}{\alpha }+\frac{%
A_{r}^{2}e^{4\gamma }}{4r^{2}}+\frac{A_{t}^{2}e^{4\gamma }}{4\alpha r^{2}}%
+e^{2\left( \eta -\gamma \right) }\rho ,  \label{c2} \\
\frac{\alpha _{r}}{2r}& =\alpha e^{2\left( \eta -\gamma \right) }\left(
P_{1}-\rho \right) .  \label{c3}
\end{align}

\underline{The Einstein-matter evolution equations:}

\begin{eqnarray}
\eta _{tt}-\alpha \eta _{rr} &=&\frac{\alpha _{rr}}{2}+\frac{\alpha _{t}\eta
_{t}}{2\alpha }-\frac{\alpha _{r}^{2}}{4\alpha }+\frac{\alpha _{r}\eta _{r}}{%
2}+\alpha \gamma _{r}^{2}-\gamma _{t}^{2}+  \label{e1} \\
&&+\frac{e^{4\gamma }}{4r^{2}}\left( A_{t}^{2}-\alpha A_{r}^{2}\right)
+\alpha e^{2\left( \eta -\gamma \right) }\left( P_{2}-P_{3}\right) ,  \notag
\end{eqnarray}%
\begin{align}
\gamma _{tt}-\alpha \gamma _{rr}& =\frac{\alpha _{t}\gamma _{t}}{2\alpha }+%
\frac{\alpha \gamma _{r}}{r}+\frac{\alpha _{r}\gamma _{r}}{2}+\frac{%
e^{4\gamma }}{2r^{2}}\left( A_{t}^{2}-\alpha A_{r}^{2}\right) +  \label{e2}
\\
& +\frac{\alpha e^{2\left( \eta -\gamma \right) }}{2}\left( \rho
-P_{1}+P_{2}-P_{3}\right) ,  \notag \\
A_{tt}-\alpha A_{rr}& =\frac{A_{t}\alpha _{t}}{2\alpha }+\frac{A_{r}\alpha
_{r}}{2}-4A_{t}\gamma _{t}+4\alpha A_{r}\gamma _{r}+  \label{e3} \\
& -\frac{\alpha A_{r}}{r}+2\alpha re^{2\eta -4\gamma }S_{23}.  \notag
\end{align}

\underline{The Vlasov equation:}%
\begin{align}
& \frac{\partial f}{\partial t}+\frac{\sqrt{\alpha }v^{1}}{v^{0}}\frac{%
\partial f}{\partial r}-\left[ \left( \sqrt{\alpha }\left( \eta _{r}-\gamma
_{r}\right) +\frac{\alpha _{r}}{2\sqrt{\alpha }}\right) v^{0}\right] \frac{%
\partial f}{\partial v^{1}}+  \notag \\
& +\left[ \sqrt{\alpha }\gamma _{r}\frac{\left( v^{2}\right) ^{2}}{v^{0}}-%
\sqrt{\alpha }\left( \gamma _{r}-\frac{1}{r}\right) \frac{\left(
v^{3}\right) ^{2}}{v^{0}}+\frac{\sqrt{\alpha }A_{r}e^{2\gamma }}{r}\frac{%
v^{2}v^{3}}{v^{0}}\right] \frac{\partial f}{\partial v^{1}}+  \notag \\
& -\left( \eta _{t}-\gamma _{t}\right) v^{1}\frac{\partial f}{\partial v^{1}}%
-\left[ \gamma _{t}v^{2}+\sqrt{\alpha }\gamma _{r}\frac{v^{1}v^{2}}{v^{0}}%
\right] \frac{\partial f}{\partial v^{2}}+  \label{v} \\
& -\left[ \frac{\sqrt{\alpha }A_{r}e^{2\gamma }}{r}\frac{v^{1}v^{2}}{v^{0}}+%
\sqrt{\alpha }\left( \frac{1}{r}-\gamma _{r}\right) \frac{v^{1}v^{3}}{v^{0}}%
\right] \frac{\partial f}{\partial v^{3}}+  \notag \\
& +\left[ -\frac{A_{t}e^{2\gamma }}{r}v^{2}+\gamma _{t}v^{3}\right] \frac{%
\partial f}{\partial v^{3}}=0,  \notag
\end{align}%
where the variables $v^{a}$ are related to the canonical momentum variables $%
p^{a}$ by%
\begin{equation*}
v^{0}=\sqrt{\alpha }e^{\eta -\gamma }p^{0},\text{ }v^{1}=e^{\eta -\gamma
}p^{1},\text{ }v^{2}=e^{\gamma }p^{2}+Ae^{\gamma }p^{3},\text{ }%
v^{3}=re^{-\gamma }p^{3},
\end{equation*}

and the matter terms are defined by%
\begin{align*}
\rho & =\int_{R^{3}}f(t,r,v)v^{0}dv, \\
J& =\int_{R^{3}}f(t,r,v)v^{1}dv, \\
P_{k}& =\int_{R^{3}}f(t,r,v)\frac{\left( v^{k}\right) ^{2}}{v^{0}}dv,\text{ }%
k=1,2,3, \\
S_{23}& =\int_{R^{3}}f(t,r,v)\frac{v^{2}v^{3}}{v^{0}}dv.
\end{align*}%

We assume that all particles have proper mass one, so 
\begin{equation*}
g_{ab}p^{a}p^{b}=-1,
\end{equation*}%
and %
\begin{equation*}
v^{0}=\sqrt{1+\left( v^{1}\right) ^{2}+\left( v^{2}\right) ^{2}+\left(
v^{3}\right) ^{2}}.
\end{equation*}%

Observe that $f$ does not depend on $z$ and $\theta $ because of the imposed
symmetry assumptions. We assume initial data $\alpha_{0}=\alpha(0,r)$, $\gamma_{0}=\gamma(0,r)$, $\gamma_{1}=\gamma_{t}(0,r)$, $\eta_{0}=\eta(0,r)$, $\eta_{1}=\eta_{t}(0,r)$, $A_{0}=A(0,r)$, $A_{1}=A_{t}(0,r)$ and $f_{0}=f(0,r,v)$ which is smooth and furthermore, according to the definition of asymptotic flatness we say that the initial data is asymptotically flat if there exists a set $K$ which is compact in $\{\vec{x}\in \mathbb{R}^{3}:z=const.\}$, such that $f_{0}\equiv 0$ in the complement of $K$ and $\gamma_{0},$ $\eta_{0}$ and $A_{0}$ are constant in the complement of $K.$ By a rescaling of $z$ we can obtain $\gamma_{0}=0$ in the complement of $K.$ It is important to point out that not only a plane is flat in two dimensions, also a cone is flat in two dimensions and the asymptotic behavior will in general be conical which admit $\eta_{0}$ and $A_{0}$ to be different from zero in the complement of $K$. Furthermore $\gamma_{1}\equiv\eta_{1}\equiv A_{1}\equiv 0$ in the complement of $K$. For the function $\alpha_{0} $ we impose that $\underset{%
r\rightarrow \infty }{\lim }\alpha_{0}(r) \equiv C_{\alpha },$ for some
constant $C_{\alpha }$. A natural choice is $C_{\alpha }=1$. See \cite{BCM} for more details concerning the asymptotic flatness
conditions.  

Regularity of the metric lead to boundary conditions of the metric on the axis of symmetry. In \cite{C} the following lemma is proved.

\begin{lemma}
Let $A_{\mu\nu}$ be a smooth tensor field on the open ball $B\left( \epsilon
\right) \subset\mathbb{R}^{3},$ $\epsilon >0,$ $A_{\mu\nu}$ invariant under rotation
around the z-axis. Then there exist smooth functions $\chi $, $\varphi $, $%
\psi $, $\sigma $, $\varrho $ and $\lambda $ invariant under rotations
around the z-axis, such that $A_{\mu\nu}$ takes the form $\left( r,\theta ,z-%
\text{cylindrical coordinates}\right) $%
\begin{equation*}
A_{\mu\nu}dx^{\mu}dx^{\nu}=\chi dz^{2}+\varphi r^{2} dzd\theta +\psi rdzdr+\sigma
r^{2}dr^{2}+\varrho\left( dr^{2}+r^{2}d\theta ^{2}\right) +\lambda r^{3}drd\theta .
\end{equation*}
\end{lemma}

For a proof, see \cite[appendix C]{C}. In our case this in particular
implies 
\begin{align*}
A_{0}(0)& =O(r^{2}), r\rightarrow 0,\\
\eta_{0}(0)& =O(1), r\rightarrow 0.
\end{align*}%
Furthermore the ratio between the circumference
of a small circle around $r=0$ and its radius must tend to $2\pi $ as $%
r\rightarrow 0$. This and the fact that $A_{0}(0)=O(r^{2})$ implies that $\eta_{0}(0)=0$. Hence%
\begin{align}
A_{0}(0)& =O(r^{2}), r\rightarrow 0, \\
\eta_{0}(0)& =0, r\rightarrow 0. 
\end{align}
Hence we call the initial data smooth, cylindrically symmetric and asymptotically flat if it is $C^{\infty}$ and satisfy the boundary conditions mentioned above, see equations $\left (\ref{b5}\right )-\left (\ref{a2}\right )$ with $t=0$.

From \cite[p 90-91]
{FR}, we have that the symmetry assumptions on the initial surface are
preserved by the time evolution. We will consider the solution given by Choquet-Bruhat \cite{CB} which exists and is regular on some time interval $[0,T).$ Since $\alpha $ is bounded on any closed subinterval and the characteristic system associated with the Vlasov equation, see $\left( \ref{mat1}\right) $ and $\left( \ref{mat}\right) $, yields that the matter is compactly supported on the same subinterval, it follows from $\left( \ref{e1} \right) $-$\left( \ref{e3}\right) $ that the asymptotic flatness conditions hold for $\gamma ,$ $\eta ,$ and $A$ due to the hyperbolic nature of these equations. Hence the boundary conditions are%
\begin{equation}
\underset{r\rightarrow \infty }{\lim }f(t,r,v)=0, \label{b5} 
\end{equation}
\begin{equation}
\underset{r\rightarrow \infty }{\lim }\gamma (t,r)=0,  \label{b1} 
\end{equation}
\begin{equation}
\underset{r\rightarrow \infty }{\lim }\eta (t,r)=C_{\eta },  \label{b2} 
\end{equation}
\begin{equation}
\underset{r\rightarrow \infty }{\lim }A(t,r)=C_{A},  \label{b3}
\end{equation}
\begin{equation}
\underset{r\rightarrow \infty }{\lim }\alpha (t,r)=1,  \label{b4} 
\end{equation}
\begin{equation}
A(t,r)=O(r^{2}), r\rightarrow 0, \label{a1} 
\end{equation}
\begin{equation}
\eta(t,r)=0, r\rightarrow 0,  \label{a2}
\end{equation}

where the solution exists. To conclude this section we mention a lemma for cylindrically symmetric, asymptotically flat spacetimes, which show that these spacetimes are well-behaved in a certain sense. Namely, the Penrose singularity theorem \cite[chapter 9]{Wald} says that the presence of a trapped surface implies the
existence of a singularity, but the following lemma shows that trapped surfaces are absent.

\begin{lemma}
Asymptotically flat, cylindrically symmetric initial data ($S,g_{\alpha
\beta },K_{\alpha \beta }),$ with matter satisfying a reasonable energy
condition, do not contain trapped surfaces which are either compact or
invariant under the isometry group.
\end{lemma}

For a proof and to see what is meant by a reasonable energy condition, see \cite{BCM}. The important thing for us is that Vlasov matter satisfies this condition. Thus possible singularities will not form due to the presence of trapped surfaces.

\section{The main theorems}

We are going to prove that if singularities form in the evolution, the first one occurs at the axis of symmetry. To obtain this result we will assume that the derivatives of the metric functions and the matter distribution function are well-behaved at the axis of symmetry so that the difficulties that arise there are eliminated. In the polarized case, i.e. when $A\equiv 0$ however, we will not need any assumption on the metric due to the following theorem.

\begin{theorem} 
Given $\epsilon >0$, let $(\mathbb{R}^{3},h_{\mu\nu},K_{\mu\nu},f_{0})$ be smooth, asymptotically flat, cylindrically symmetric initial data which satisfy the constraint equations and where the metric is coordinatized as $ds^{2}=-\alpha e^{2(\eta-\gamma)}dt^2+e^{2(\eta-\gamma)}dr^2+e^{2\gamma}dz^{2}+r^{2}e^{-2\gamma}d\theta^{2}$. Let $(f,g_{ab})$ be the local smooth solution associated to the initial data and let $[0,T)$ be the maximal time interval of existence. Assume that for $r<\epsilon$
\begin{equation}
 f(t,r,v)\equiv 0, \forall t\in [0,T), \forall v\in\mathbb{R}^{3}. \label{a}
\end{equation}

Then $T=\infty$.

\end{theorem}

Due to the fact that the matter is supported away from the axis of symmetry one may suspect that we could use the vacuum results of \cite{BCM} to control the field at the axis of symmetry. However this does not seem straightforward, due to the fact that $\alpha$ and $\alpha_{t}$ will be functions of $t$ when $r<\epsilon$. Hence we can not reduce the nonpolarized Einstein equations to a wavemap from $2+1$-dimensional Minkowski space to the hyperbolic plane with the metric $h_{\mu\nu}dx^{\mu}dx^{\nu}=d\gamma^{2}+\frac{e^{-4\gamma}}{4}d\omega^{2}$ in a vicinity of the axis of symmetry where the matter vanishes and then apply the methods of Christodoulou/Tahvildar-Zadeh. However as the theorem above shows we are able to remove the assumptions on the derivatives of the metric in the polarized case, i.e. $A\equiv 0$. This is done by a combination of Sobolev estimates and the arguments in the proof of the following theorem.

\begin{theorem} 
Given $\epsilon >0,$ let $\left ( \mathbb{R}^{3},h_{\mu\nu },K_{\mu\nu},f_{0}\right ) $ be smooth, asymptotically flat, cylindrically symmetric initial data which satisfy the constraint equations and where the metric is coordinatized as in $\left ( \ref{met}\right ) .$ Let $\left ( f,g_{ab}\right ) $ be the local smooth solution associated to the initial data and let $[0,T)$ be the maximal time interval of existence. Assume that there exists a constant $C_{\epsilon }$ such that for $r<\epsilon $

\begin{equation}
\left| \partial ^{\beta }\gamma \right| \leq C_{\epsilon },
\text{ and }\left| \partial ^{\beta }A\right| \leq C_{\epsilon }r,\text{ }
\forall \beta \in\mathbb{N}, \forall t\in [0,T),\text{\label{as1}} 
\end{equation}
\begin{equation}
f(t,r,v)\equiv 0,\forall t\in [0,T).  \label{as2}
\end{equation}

Then $T=\infty $.

\end{theorem}

In assumption $\left( \ref{as1}\right) $ above we used the notation that $%
\beta =(\beta _{1},\beta _{2})$ is a multi-index and $\partial ^{\beta
}=\partial _{t}^{\beta _{1}}\partial _{r}^{\beta _{2}}.$

\begin{remark}By assuming that the support of the $v^{2}$ and $v^{3}$ momentum is uniformly bounded in time and that the modulus of ``angular momentum'' (which is more or less a product between the radial position of the particle and the velocity in $\theta$-direction) and the velocity in the $z$-direction is bounded away from zero initially global existence is an immediate consequence of Theorem 1. This can be seen by using the Killing equations $\left (\ref{K1}\right )$ and $\left (\ref{K2}\right )$ together with the assumptions above to conclude that the matter never reaches $r=0$ in finite time. 
\end{remark}

\section{Proof of theorem 2}

Given $\epsilon>0$. If $T=\infty$ we are done so assume that $T<\infty$. If we are able to obtain uniform bounds on all the functions (including their derivatives) we can extend the solution $(f,g_{ab})$ to $t=T$, and then apply the local existence theorem again with initial data at $t=T$, contradicting the fact that $[0,T)$ was the maximal time of existence. The outline of the proof follows \cite{And1} and we briefly summarize the strategy of the proof. In step 1 we bound the metric functions by using a conserved ''energy''. In step 2 we bound $Q(t)$ and the derivatives of the metric functions by using a light cone argument together with a crucial lemma, (Lemma 3). An important ingredient in the proof of this lemma is a rescaling of the momentum variables leading to cancellation of
terms that are difficult to handle. In step 3 we bound the derivatives of the matter terms. Here the geodesic deviation equation plays an important role. It is then straightforward to see that the higher order derivatives can be bounded by similar arguments as will be outlined in step 4. Whenever convenient we shall use the convention to write $C$ for arbitrary constants even if they change from line to line. 

\underline{$Step\text{ } 1:$} (Bounds on the metric functions.)

The asymptotic flatness condition $\left( \ref{b2}\right) $ is crucial and
leads to a uniform bound of $\eta $ and $\alpha ,$ because $\eta _{r}\geq 0.$
Indeed by $\left( \ref{c2}\right) $ and $\left (\ref{a2}\right )$

\begin{equation*}
\left| \eta (t,r)\right| =\left| \int_{0}^{r}\eta _{r}(t,r^{\prime
})dr^{\prime }\right| \leq \left| \int_{0}^{\infty }\eta _{r}(t,r)dr\right|
=\underset{r\rightarrow\infty}{\lim}\left |\eta(t,r)\right |=C_{\eta }.
\end{equation*}

Hence $\left| \eta \right| $ is uniformly bounded on $[0,T)\times \lbrack
0,\infty ).$ Note that the last equality above can be written%
\begin{equation}
\int_{0}^{\infty }\eta _{r}(t,r)dr=\int_{0}^{\infty }r\left[ \gamma _{r}^{2}+%
\frac{\gamma _{t}^{2}}{\alpha }+\frac{A_{r}^{2}e^{4\gamma }}{4r^{2}}+\frac{%
A_{t}^{2}e^{4\gamma }}{4\alpha r^{2}}+e^{2\left( \eta -\gamma \right) }\rho %
\right] dr=C_{\eta }.  \label{energy}
\end{equation}

By the constraint equation $\left( \ref{c3}\right) $, equation $\left( %
\ref{energy}\right) $ and that $P_{1}\leq \rho $ 
 
\begin{align*}
\log \alpha (t,r)& =-\int_{r}^{\infty }\frac{\partial }{\partial r}\log
\alpha (t,r^{\prime })dr^{\prime }=-2\int_{r}^{\infty }r^{\prime }e^{2\left(
\eta -\gamma \right) }\left( P_{1}-\rho \right) dr^{\prime }\leq \\
& \leq 2\int_{r}^{\infty }r^{\prime }e^{2\left( \eta -\gamma \right) }\rho
dr^{\prime }\leq 2\int_{0}^{\infty }re^{2\left( \eta -\gamma \right) }\rho
dr\leq 2C_{\eta }.
\end{align*}

By the constraint equation $\left( \ref{c3}\right) $ $\alpha _{r}\leq 0$. So $\underset{r\rightarrow \infty }{\lim }\alpha (t,r)=1,$ implies that

\begin{equation}
1\leq \alpha \leq e^{2C_{\eta }}=:\hat{C}.  \label{alf}
\end{equation}

Hence $\alpha $ is uniformly bounded on $[0,T)\times \lbrack 0,\infty ).$

By the asymptotic flatness condition $\left( \ref{b1}\right) $ and the fact that $\left| \gamma (0,r)\right| =0$ in the complement of the set $K$, see $\left (\ref{b1}\right )$, we can choose $\widetilde{R}\in (0,\infty ),$ big enough such that $\left|\gamma (t,r)\right| =0,$ $\forall (t,r)\in \lbrack 0,T)\times \lbrack \widetilde{R}+\hat{C}t,\infty )$. So by H\"{o}lder's inequality
and equation $\left( \ref{energy}\right) $ we have for $r\geq \epsilon $

\begin{align*}
\left| \gamma (t,r)\right| & =\left| \int_{r}^{\widetilde{R}+\hat{C}t}\gamma
_{r}(t,r^{\prime })dr^{\prime }\right| \leq \\
& \leq \left( \int_{r}^{\widetilde{R}+\hat{C}t}\frac{1}{r^{\prime }}%
dr^{\prime }\right) ^{\frac{1}{2}}\left( \int_{r}^{\widetilde{R}+\hat{C}%
t}r^{\prime }\gamma _{r}^{2}dr^{\prime}\right) ^{\frac{1}{2}}\leq C_{\eta }\left(
\int_{r}^{\widetilde{R}+\hat{C}t}\frac{1}{r^{\prime }}dr^{\prime }\right) ^{%
\frac{1}{2}}\leq \\
& \leq C_{\eta }\left( \log \frac{\widetilde{R}+\hat{C}t}{r}\right) ^{\frac{1%
}{2}}\leq C_{\eta }\left( \log \frac{\widetilde{R}+\hat{C}T}{\epsilon }%
\right) ^{\frac{1}{2}}.
\end{align*}%
Hence $\gamma $ is uniformly bounded on $[0,T)\times \lbrack \epsilon
,\infty )$. If $r<\epsilon $ then by H\"{o}lder's inequality and assumption $%
\left( \ref{as1}\right) $%
\begin{equation*}
\left| \gamma (t,\epsilon )-\gamma (t,r)\right| =\left| \int_{r}^{\epsilon
}\gamma _{r}(t,r^{\prime })dr^{\prime }\right| \leq\epsilon\underset{r<\epsilon}{\sup }\left|\gamma_{r}(t,\cdot)\right|\leq C(\epsilon)\epsilon.
\end{equation*}

Because $\gamma (t,\epsilon )$ is bounded due to the estimate above, $%
\gamma $ is uniformly bounded on $[0,T)\times \lbrack 0,\epsilon )$. Hence $%
\gamma $ is uniformly bounded on $[0,T)\times \lbrack 0,\infty )$.

Let $\widetilde{R}\in (0,\infty )$ be as above, then $\left| A\left(
t,r\right) \right| =C_{A}$, $\forall (t,r)\in \lbrack 0,T)\times \lbrack 
\widetilde{R}+\hat{C}t,\infty ),$ which follows from the asymptotic flatness
condition $\left( \ref{b3}\right) $ and that $\left| A(0,r)\right| =C_{A}$
in the complement of the set $K,$ see $\left (\ref{b3}\right )$. So by H\"{o}lder's
inequality and equation $\left( \ref{energy}\right) $ 

\begin{equation*}
\left| C_{A}-A\left( t,r\right) \right|=\left| \int_{r}^{\widetilde{R}+
\hat{C}t}A_{r}\left( t,r^{\prime}\right) dr^{\prime }\right| \leq
\end{equation*}
\begin{equation*}
\leq \left( \int_{r}^{\widetilde{R}+\hat{C}t}4r^{\prime }e^{-4\gamma
}dr^{\prime }\right) ^{\frac{1}{2}}\left( \int_{r}^{\widetilde{R}+\hat{C}t}
\frac{A_{r}^{2}e^{4\gamma }}{4r^{\prime }}dr^{\prime }\right) ^{\frac{1}{2}
}\leq
\end{equation*}
\begin{equation*}
\leq C_{\eta }\left( \int_{r}^{\widetilde{R}+\hat{C}t}4r^{\prime
}e^{-4\gamma }dr^{\prime }\right) ^{\frac{1}{2}}\leq
\end{equation*}
\begin{equation*}
\leq \widetilde{C}(\epsilon)(\widetilde{R}+\hat{C}T).
\end{equation*}

Hence $A$ is uniformly bounded on $[0,T)\times \lbrack 0,\infty )$.

\bigskip

\underline{$Step\text{ } 2:$} (Bounds on the first order derivatives of the metric and $Q(t).$)

Define the support of the momentum $Q(t)$ by
\begin{equation}
Q(t):=\sup \{\left| v\right|+3 :\exists (s,r)\in \lbrack 0,t]\times \lbrack
0,\infty )\text{ such that }f(s,r,v)\neq 0\}, \label{Q}
\end{equation}%
and define the two quadratic forms $G$ and $H$ by%
\begin{equation*}
G:=\frac{1}{2}(\frac{\gamma _{t}^{2}}{\alpha }+\gamma _{r}^{2})+\frac{%
e^{4\gamma }}{8r^{2}}(\frac{A_{t}^{2}}{\alpha }+A_{r}^{2}),
\end{equation*}%
\begin{equation*}
H:=\frac{\gamma _{t}\gamma _{r}}{\sqrt{\alpha }}+\frac{e^{4\gamma }}{4\sqrt{%
\alpha }r^{2}}A_{t}A_{r}.
\end{equation*}%
The aim is to show that the quantity%
\begin{equation*}
\Gamma (t):=\underset{r\in \lbrack\epsilon,\infty )}{\sup G(t},%
\text{%
$\cdot$%
})+Q^{2}(t),
\end{equation*}%
is uniformly bounded on $[0,T),$ by obtaining a Gr\"{o}nvall type of
inequality%
\begin{equation*}
\Gamma (t)\leq C_{1}+C_{2}\int_{t_{0}}^{t}\Gamma (s)\log \Gamma (s)ds.
\end{equation*}

First of all, if $\phi $ is a geodesic and $X$ is a Killing vector field,
then $g(X,\phi ^{\prime })$ is conserved along the geodesic, see \cite[p
442]{Wald}. The particles follow the geodesics of spacetime with tangent 
$p^{\mu },$ so $g_{\mu \nu }p^{\mu }\left( \partial _{z}\right) ^{\nu }$ and 
$g_{\mu \nu }p^{\mu }\left( \partial _{\theta }\right) ^{\nu }$ are
conserved. Hence

\begin{align}
g_{zz}p^{z}\left( \partial _{z}\right) ^{z}+g_{z\theta }p^{\theta }\left(
\partial _{z}\right) ^{z}& =V^{2}(t)e^{\gamma \left( t,R\left( t\right)
\right) }=C,\label{K1} \\
g_{z\theta }p^{z}\left( \partial _{\theta }\right) ^{\theta }+g_{\theta
\theta }p^{\theta }\left( \partial _{\theta }\right) ^{\theta }&
=V^{2}(t)Ae^{\gamma \left( t,R\left( t\right) \right)
}+V^{3}(t)R(t)e^{-\gamma \left( t,R\left( t\right) \right) }=C\label{K2}.
\end{align}

Here $R(t),V^{2}(t)$ and $V^{3}(t)$ are solutions to the characteristic
system associated to the Vlasov equation.
The characteristic equation for $R(t)$ is%
\begin{equation}
\frac{dR}{ds}=\frac{\sqrt{\alpha }V^{1}}{V^{0}},  \label{mat1}
\end{equation}%
and we have especially that 
\begin{equation}
\left| \frac{dR}{ds}\right| =\left| \frac{\sqrt{\alpha }V^{1}}{V^{0}}\right|
\leq \sqrt{\alpha }\leq \sqrt{\hat{C}}\text{.\label{mat}}
\end{equation}%
So $R(t)$ is uniformly bounded on $[0,T)$. More important though is that due
to assumption $\left( \ref{as2}\right) $ and that $%
f_{0}(r,v)=f(s,R(s,0,r,v),V(s,0,r,v))$ it follows that%
\begin{equation}
R(t)\geq \epsilon >0,\forall t\in \lbrack 0,T).  \label{ax}
\end{equation}%
Furthermore the bounds on $\left |\gamma\right |$ and $\left |A\right |$, equation $\left (\ref{ax}\right )$ and that $f_{0}$ is compactly supported implies that $\left| V^{2}(t)\right| $ and $\left| V^{3}(t)\right| $ are uniformly bounded
on $[0,T)$ in view of equations $\left (\ref{K1}\right )$ and $\left (\ref{K2}\right )$. So we can conclude that $\sup \{\left| v^{2}\right| +\left| v^{3}\right| +3:\exists \left( s,r ,v^{1}\right) \in \lbrack 0,t]\times \lbrack 0,\infty )\times\mathbb{R}$ with $f(s,r,v)\neq 0\}$ is also uniformly bounded on $[0,T).$
Hence in order to control $Q(t)$ it is sufficient to control $Q^{1}(t):=\sup
\{\left| v^{1}\right|+3 :\exists \left( s,r,v^{2},v^{3}\right) \in \lbrack 0,t]\times\lbrack 0,\infty )\times\mathbb{R}^{2}$ with $f(s,r,v)\neq 0\}.$
We also notice a few facts about the matter terms which follow from the
discussion above. In the $\rho -P_{1}$ term some cancellation occurs which
is useful to take advantage of. Indeed we have%
\begin{align}
0& \leq \left( \rho -P_{1}\right) \left( t,r\right) =\int_{R^{3}}\left(
v^{0}-\frac{\left( v^{1}\right) ^{2}}{v^{0}}\right) f(t,r,v)dv=  \notag \\
& =\int_{R^{3}}\left( \frac{1+\left( v^{2}\right) ^{2}+\left( v^{3}\right)
^{2}}{v^{0}}\right) f(t,r,v)dv\leq  \notag \\
& \leq \int_{R^{3}}\left( 1+\left( v^{2}\right) ^{2}+\left( v^{3}\right)
^{2}\right) \left| f\right| \frac{dv}{\sqrt{1+\left( v^{1}\right) ^{2}}}\leq
\label{ma1} \\
& \leq C\left\| f_{0}\right\| \int_{\left| v^{1}\right| \leq Q^{1}(t)}\frac{%
dv^{1}}{\sqrt{1+\left( v^{1}\right) ^{2}}}\leq  \notag \\
& \leq C\log Q^{1}(t).  \notag
\end{align}%
In a similar way the following estimates for $P_{k}(t,r)$ $,$ $k=2,3$ hold%
\begin{align}
0& \leq P_{k}(t,r)=\int_{R^{3}}\frac{\left( v^{k}\right) ^{2}}{v^{0}}%
f(t,r,v)dv\leq  \notag \\
& \leq C\left\| f_{0}\right\| \int_{\left| v^{1}\right| \leq Q^{1}(t)}\frac{%
dv^{1}}{\sqrt{1+\left( v^{1}\right) ^{2}}}\leq  \label{m2} \\
& \leq C\log Q^{1}(t).  \notag
\end{align}%
The argument for $S_{23}$ is similar, hence%
\begin{equation}
S_{23}\leq C\log Q^{1}(t).  \label{m3}
\end{equation}%

Next, define two vector fields by

\begin{align*}
\xi :=\frac{1}{\sqrt{2}}(\partial _{t}-\sqrt{\alpha }\partial _{r}), \\
\zeta :=\frac{1}{\sqrt{2}}(\partial _{t}+\sqrt{\alpha }\partial _{r}).
\end{align*}

These vector fields are null vector fields, i.e. $g_{ab}\xi^{a}\xi^{b}=g_{ab}\zeta^{a}\zeta^{b}=0$. Now by the evolution equations $\left( \ref{e2}\right) $ and $\left( \ref{e3}%
\right) $ 

\begin{align}
\xi \left( G+H\right) & =\frac{1}{\sqrt{2}}(\partial _{t}-\sqrt{\alpha }%
\partial _{r})\left[ \frac{1}{2}(\frac{\gamma _{t}}{\sqrt{\alpha }}+\gamma
_{r})^{2}+\frac{e^{4\gamma }}{8r^{2}}(\frac{A_{t}}{\sqrt{\alpha }}+A_{r})^{2}%
\right] =  \notag \\
& =\frac{1}{\sqrt{2}}(\frac{\gamma _{t}}{\sqrt{\alpha }}+\gamma _{r})\left( 
\frac{\gamma _{tt}}{\sqrt{\alpha }}-\frac{\gamma _{t}\alpha _{t}}{2\alpha ^{%
\frac{3}{2}}}+\gamma _{tr}-\gamma _{tr}+\frac{\gamma _{t}\alpha _{r}}{%
2\alpha }-\sqrt{\alpha }\gamma _{rr}\right) +  \notag \\
& +\frac{e^{4\gamma }}{8r^{2}}\frac{2}{\sqrt{2}}(\frac{A_{t}}{\sqrt{\alpha }}%
+A_{r})\left( \frac{A_{tt}}{\sqrt{\alpha }}-\frac{A_{t}\alpha _{t}}{2\alpha
^{\frac{3}{2}}}+A_{tr}-A_{tr}\right) +  \notag \\
& +\frac{e^{4\gamma }}{8r^{2}}\frac{2}{\sqrt{2}}(\frac{A_{t}}{\sqrt{\alpha }}%
+A_{r})\left( \frac{A_{t}\alpha _{r}}{2\alpha }-\sqrt{\alpha }A_{rr}\right) +
\notag \\
& +\frac{e^{4\gamma }}{8r^{2}}\frac{1}{\sqrt{2}}(\frac{A_{t}}{\sqrt{\alpha }}%
+A_{r})^{2}\left( 4\gamma _{t}-4\sqrt{\alpha }\gamma _{r}+\frac{2\sqrt{%
\alpha }}{r}\right)=  \label{tao} \\
& =\frac{1}{\sqrt{2}}(\frac{\gamma _{t}}{\sqrt{\alpha }}+\gamma _{r})\left( 
\frac{\gamma _{t}\alpha _{r}}{2\alpha }+\frac{\alpha _{r}\gamma _{r}}{2\sqrt{%
\alpha }}+\frac{\sqrt{\alpha }\gamma _{r}}{r}+\frac{\sqrt{\alpha }e^{2\left(
\eta -\gamma \right) }}{2}\kappa \right) +  \notag \\
& +\frac{e^{4\gamma }}{8r^{2}}\frac{1}{\sqrt{2}}\left( \frac{A_{t}^{2}\alpha
_{r}}{\alpha ^{\frac{3}{2}}}+\frac{2A_{r}A_{t}\alpha _{r}}{\alpha }+\frac{%
2A_{t}^{2}}{\sqrt{\alpha }r}+4rA_{t}e^{2\eta -4\gamma }S_{23}\right) + 
\notag \\
& +\frac{e^{4\gamma }}{8r^{2}}\frac{1}{\sqrt{2}}\left( \frac{A_{r}^{2}\alpha
_{r}}{\sqrt{\alpha }}+\frac{2A_{t}A_{r}}{r}+4\sqrt{\alpha }rA_{r}e^{2\eta
-4\gamma }S_{23}\right)=  \notag \\
& =\frac{\alpha _{r}}{\sqrt{2\alpha }}\left( G+H\right) +\frac{\kappa
e^{2\left( \eta -\gamma \right) }}{2}\zeta \left( \gamma \right) +\frac{%
\gamma _{r}}{r}\zeta \left( \gamma \right) +  \notag \\
& +\frac{S_{23}e^{2\eta }}{2r}\zeta \left( A\right) +\frac{A_{t}e^{4\gamma }%
}{4\sqrt{\alpha }r^{3}}\zeta \left( A\right) ,  \notag
\end{align}%

and%
\begin{equation*}
\zeta \left( G-H\right )=\frac{1}{\sqrt{2}}(\partial _{t}+\sqrt{\alpha }%
\partial _{r})\left[ \frac{1}{2}(\frac{\gamma _{t}}{\sqrt{\alpha }}-\gamma
_{r})^{2}+\frac{e^{4\gamma }}{8r^{2}}(\frac{A_{t}}{\sqrt{\alpha }}-A_{r})^{2}%
\right] =
\end{equation*}

\begin{align}
& =\frac{1}{\sqrt{2}}(\frac{\gamma _{t}}{\sqrt{\alpha }}-\gamma _{r})\left( 
\frac{\gamma _{tt}}{\sqrt{\alpha }}-\frac{\gamma _{t}\alpha _{t}}{2\alpha ^{%
\frac{3}{2}}}+\gamma _{tr}-\gamma _{tr}-\frac{\gamma _{t}\alpha _{r}}{%
2\alpha }-\sqrt{\alpha }\gamma _{rr}\right) + \notag \\
& +\frac{e^{4\gamma }}{8r^{2}}\frac{2}{\sqrt{2}}(\frac{A_{t}}{\sqrt{\alpha }}%
-A_{r})\left( \frac{A_{tt}}{\sqrt{\alpha }}-\frac{A_{t}\alpha _{t}}{2\alpha
^{\frac{3}{2}}}+A_{tr}-A_{tr}\right) + \notag \\
& +\frac{e^{4\gamma }}{8r^{2}}\frac{2}{\sqrt{2}}(\frac{A_{t}}{\sqrt{\alpha }}%
-A_{r})\left( \frac{A_{t}\alpha _{r}}{2\alpha }-\sqrt{\alpha }A_{rr}\right) +
\notag \\  
& +\frac{e^{4\gamma }}{8r^{2}}\frac{1}{\sqrt{2}}(\frac{A_{t}}{\sqrt{\alpha }}%
-A_{r})^{2}\left( 4\gamma _{t}+4\sqrt{\alpha }\gamma _{r}-\frac{2\sqrt{%
\alpha }}{r}\right)= \label{zeta} \\
& =\frac{1}{\sqrt{2}}(\frac{\gamma _{t}}{\sqrt{\alpha }}-\gamma _{r})\left( -%
\frac{\gamma _{t}\alpha _{r}}{2\alpha }+\frac{\alpha _{r}\gamma _{r}}{2\sqrt{%
\alpha }}+\frac{\sqrt{\alpha }\gamma _{r}}{r}+\frac{\sqrt{\alpha }e^{2\left(
\eta -\gamma \right) }}{2}\kappa \right) + \notag \\
& +\frac{e^{4\gamma }}{8r^{2}}\frac{1}{\sqrt{2}}\left( -\frac{%
A_{t}^{2}\alpha _{r}}{\alpha ^{\frac{3}{2}}}+\frac{2A_{r}A_{t}\alpha _{r}}{%
\alpha }-\frac{2A_{t}^{2}}{\sqrt{\alpha }r}+4rA_{t}e^{2\eta -4\gamma
}S_{23}\right) +  \notag \\
& +\frac{e^{4\gamma }}{8r^{2}}\frac{1}{\sqrt{2}}\left( -\frac{%
A_{r}^{2}\alpha _{r}}{\sqrt{\alpha }}+\frac{2A_{t}A_{r}}{r}-4\sqrt{\alpha }%
rA_{r}e^{2\eta -4\gamma }S_{23}\right)= \notag \\
& =\frac{-\alpha _{r}}{\sqrt{2\alpha }}\left( G-H\right) +\frac{\kappa
e^{2\left( \eta -\gamma \right) }}{2}\xi \left( \gamma \right) +\frac{%
\gamma _{r}}{r}\xi \left( \gamma \right)+ \notag \\
& +\frac{S_{23}e^{2\eta }}{2r}\xi \left( A\right) -\frac{A_{t}e^{4\gamma }}{%
4\sqrt{\alpha }r^{3}}\xi \left( A\right) ,  \notag
\end{align}

where $\kappa =\rho -P_{1}+P_{2}-P_{3}.$ Let $\check{r}_{\pm }(s)$ be the
unique solutions to the equations%
\begin{equation*}
\frac{d\check{r}_{\pm }(s)}{ds}=\pm \sqrt{\alpha (s,\check{r}_{\pm }(s))},%
\text{ }\check{r}_{\pm }(t)=r,
\end{equation*}%
which exist due to the theory of ODE because $\alpha $ is a smooth function
and $\alpha \geq 1$ by $\left( \ref{alf}\right) $. Note that because 
$r\geq \epsilon $ we can find a largest time $t_{0},$  
\begin{equation}
t_{0}\leq T-\frac{\epsilon }{2\sqrt{\hat{C}}}<T,  \label{t}
\end{equation}%
such that if we go backwards along the integral curve for $\zeta$ we do not come closer to the axis than $r=\frac{\epsilon }{2}.$ Observe also that we can control the different functions on the hypersurface $t=t_{0}$ due to the local existence theorem of Choquet-Bruhat \cite{CB}. Integrate the equations $\left( \ref{tao}\right) $ and $\left( \ref{zeta}\right) $ for the quadratic
forms above along the integral curves corresponding to the vector fields $%
\zeta $ and $\xi ,$ from the hypersurface $t=t_{0}<T,$ to the point $%
(t_{,}r)\in (t_{0},T)\times \lbrack \epsilon ,\infty )$ to obtain%
\begin{align*}
& \left( G+H\right) (t,r)-\left( G+H\right) (t_{0},\check{r}_{-}(t_{0}))= \\
& =\int_{t_{0}}^{t}\left( \frac{\alpha _{r}}{\sqrt{\alpha }}\left(
G+H\right) +\frac{\kappa e^{2\left( \eta -\gamma \right) }}{2}\zeta \left(
\gamma \right) +\frac{S_{23}e^{2\eta }}{2r}\zeta \left( A\right) \right)
\left( s,\check{r}_{-}(s)\right) ds+ \\
& +\int_{t_{0}}^{t}\left( \frac{\gamma _{r}}{r}\zeta \left( \gamma \right) +%
\frac{A_{t}e^{4\gamma }}{4\sqrt{\alpha }r^{3}}\zeta \left( A\right) \right)
\left( s,\check{r}_{-}(s)\right) ds,
\end{align*}%
and%
\begin{align*}
& \left( G-H\right) (t,r)-\left( G-H\right) (t_{0},\check{r}_{+}(t_{0}))= \\
& =\int_{t_{0}}^{t}\left( \frac{-\alpha _{r}}{\sqrt{\alpha }}\left(
G-H\right) +\frac{\kappa e^{2\left( \eta -\gamma \right) }}{2}\xi \left(
\gamma \right) +\frac{S_{23}e^{2\eta }}{2r}\xi \left( A\right) \right)
\left( s,\check{r}_{+}(s)\right) ds+\\
& +\int_{t_{0}}^{t}\left( \frac{\gamma _{r}}{r}\xi \left( \gamma \right) -%
\frac{A_{t}e^{4\gamma }}{4\sqrt{\alpha }r^{3}}\xi \left( A\right) \right)
\left( s,\check{r}_{+}(s)\right) ds.
\end{align*}%
Add the equations to obtain%
\begin{align*}
& G(t,r)= \\
& =\frac{1}{2}\left[ G(t_{0},\check{r}_{-}(t_{0}))+G(t_{0},\check{r}%
_{+}(t_{0}))\right] + \\
& +\frac{1}{2}\left[ H(t_{0},\check{r}_{-}(t_{0}))-H(t_{0},\check{r}%
_{+}(t_{0}))\right] + \\
& +\frac{1}{2}\int_{t_{0}}^{t}\left( \frac{\alpha _{r}}{\sqrt{\alpha }}%
\left( G+H\right) +\frac{\kappa e^{2\left( \eta -\gamma \right) }}{2}\zeta
\left( \gamma \right) +\frac{S_{23}e^{2\eta }}{2r}\zeta \left( A\right)
\right) \left( s,\check{r}_{-}(s)\right) ds+ \\
& +\frac{1}{2}\int_{t_{0}}^{t}\left( \frac{\gamma _{r}}{r}\zeta \left(
\gamma \right) +\frac{A_{t}e^{4\gamma }}{4\sqrt{\alpha }r^{3}}\zeta \left(
A\right) \right) \left( s,\check{r}_{-}(s)\right) ds+ \\
& +\frac{1}{2}\int_{t_{0}}^{t}\left( \frac{-\alpha _{r}}{\sqrt{\alpha }}%
\left( G-H\right) +\frac{\kappa e^{2\left( \eta -\gamma \right) }}{2}\xi
\left( \gamma \right) +\frac{S_{23}e^{2\eta }}{2r}\xi\left( A\right)
\right) \left( s,\check{r}_{+}(s)\right) ds+ \\
& +\frac{1}{2}\int_{t_{0}}^{t}\left( \frac{\gamma _{r}}{r}\xi \left(
\gamma \right) -\frac{A_{t}e^{4\gamma }}{4\sqrt{\alpha }r^{3}}\xi \left(
A\right) \right) \left( s,\check{r}_{+}(s)\right) ds.
\end{align*}%
Observe that $G+H$ is always nonnegative and also from the constraint equation $\left( \ref{c3}\right) $ that $\frac{\alpha _{r}}{\alpha }\leq 0,$ so the
term $\frac{\alpha _{r}}{\sqrt{\alpha }}\left( G+H\right) $ can be dropped
which simplifies the estimate but it is not necessary. Observe also that $H\leq
G$ because of the elementary inequality $2ab\leq a^{2}+b^{2}.$ Take the
supremum with respect to $r\in \lbrack \epsilon,\infty )$ and
use that the matter is compactly supported, equations $\left( \ref{ma1}\right) $-$\left( \ref{m3}\right) $, together with the constraint equation $\left (\ref{c3}\right )$, assumption $\left (\ref{as1}\right )$ and observe that the radial distance from the axis is bounded from below by $\frac{\epsilon}{2}$, then 

\begin{align*}
\underset{r\in \lbrack \epsilon,\infty )}{\sup }G\left( t,\text{%
$\cdot$%
}\right) & \leq \underset{r\in \lbrack \epsilon,\infty )}{\sup }%
G\left( t_{0},\text{%
$\cdot$%
}\right) +\underset{r\in \lbrack \epsilon,\infty )}{\sup }H\left(
t_{0},\text{%
$\cdot$%
}\right) +  \notag \\
& +\underset{r\in \lbrack \frac{\epsilon }{2},\infty )}{\sup }%
\int_{t_{0}}^{t}C\left( \frac{\kappa e^{2\left( \eta -\gamma \right) }}{2}+%
\frac{S_{23}e^{2\eta }}{2}\right) \sqrt{G(s,\text{%
$\cdot$%
})}ds+  \notag \\ 
& +\underset{r\in \lbrack \frac{\epsilon }{2},\infty )}{\sup }%
\int_{t_{0}}^{t}Cr{\sqrt{\alpha }e^{2(\eta-\gamma)}\log Q^{1}(s)\left( G(s,\text{%
$\cdot$%
})-H(s,\text{%
$\cdot$%
})\right) }ds+  
\end{align*}

\begin{align}
& +\underset{r\in \lbrack \frac{\epsilon }{2},\infty )}{\sup }\frac{1}{2}%
\int_{t_{0}}^{t}\left( \frac{\gamma _{r}}{r}\zeta \left( \gamma \right) +%
\frac{A_{t}e^{4\gamma }}{4\sqrt{\alpha }r^{3}}\zeta \left( A\right) \right)
\left( s,\text{%
$\cdot$%
}\right) ds+  \notag \\
& +\underset{r\in \lbrack \frac{\epsilon }{2},\infty )}{\sup }\frac{1}{2}%
\int_{t_{0}}^{t}\left( \frac{\gamma _{r}}{r}\xi \left( \gamma \right) -%
\frac{A_{t}e^{4\gamma }}{4\sqrt{\alpha }r^{3}}\xi \left( A\right) \right)
\left( s,\text{%
$\cdot$%
}\right) ds\leq   \label{quad1}
\end{align}

\begin{equation*}
\leq C_{1}+C_{2}(\epsilon)\int_{t_{0}}^{t}\log Q^{1}(s)\underset{r\in \lbrack 
\epsilon ,\infty )}{\sup }G(s,\text{%
$\cdot$%
})ds+C_{3}(\epsilon)\int_{t_{0}}^{t}\underset{r\in \lbrack 
\epsilon ,\infty )}{\sup }G(s,\text{%
$\cdot$%
})ds.
\end{equation*}%
Because we are interested in the case when $\underset{r\in[\epsilon,\infty)}{\sup}G$ is large, we assume throughout the paper that $\underset{r\in[\epsilon,\infty)}{\sup}\sqrt{G}\leq \underset{r\in[\epsilon,\infty)}{\sup}G.$ The following crucial lemma will yield a bound$\ $on $\left| Q^{1}(t)\right| ^{2}$ in terms of $G$.

\begin{lemma}
Let $Q^{1}$ and $G$, be defined as above, then 
\begin{equation*}
\left( Q^{1}(t)\right) ^{2}\leq C_{1}+C_{2}\int_{t_{0}}^{t}\left[ \left(
Q^{1}(s)\right) ^{2}\log Q^{1}(s)+\underset{r\in \lbrack \epsilon 
,\infty )}{\sup }G(s,\text{%
$\cdot$%
})\right] ds.
\end{equation*}
\end{lemma}

\begin{proof}

Define $U^{1}(s):=e^{\eta-\gamma}V^{1}(s).$

The characteristic equation for $V^{1}(s)$ reads%
\begin{align*}
\frac{dV^{1}}{ds}& =-\left( \sqrt{\alpha }\left( \eta _{r}-\gamma
_{r}\right) +\frac{\alpha _{r}}{2\sqrt{\alpha }}\right) V^{0}-\left( \eta
_{t}-\gamma _{t}\right) V^{1}+ \\
& +\left[ \sqrt{\alpha }\gamma _{r}\frac{\left( V^{2}\right) ^{2}}{V^{0}}-%
\sqrt{\alpha }\left( \gamma _{r}-\frac{1}{R}\right) \frac{\left(
V^{3}\right) ^{2}}{V^{0}}+\frac{\sqrt{\alpha }A_{r}e^{2\gamma }}{R}\frac{%
V^{2}V^{3}}{V^{0}}\right] .
\end{align*}

Hence 
\begin{align*}
\frac{d}{ds}(U^{1}(s))^{2}& =2U^{1}\frac{dU^{1}}{ds}= \\
& =2U^{1}e^{\eta -\gamma }\left[ (\eta _{t}-\gamma _{t})V^{1}+(\eta
_{r}-\gamma _{r})V^{1}\frac{dR}{ds}+\frac{dV^{1}}{ds}\right] = \\
& =2U^{1}e^{\eta -\gamma }\left[ (\eta _{t}-\gamma _{t})V^{1}+(\eta
_{r}-\gamma _{r})V^{1}\frac{\sqrt{\alpha }V^{1}}{V^{0}}\right] + \\
& +2U^{1}e^{\eta -\gamma }\left[ -\left( \sqrt{\alpha }\left( \eta
_{r}-\gamma _{r}\right) +\frac{\alpha _{r}}{2\sqrt{\alpha }}\right)
V^{0}-\left( \eta _{t}-\gamma _{t}\right) V^{1}\right] + \\
& +2U^{1}e^{\eta -\gamma }\left[ \sqrt{\alpha }\gamma _{r}\frac{\left(
V^{2}\right) ^{2}}{V^{0}}-\sqrt{\alpha }\left( \gamma _{r}-\frac{1}{R}%
\right) \frac{\left( V^{3}\right) ^{2}}{V^{0}}\right] + \\
& +2U^{1}e^{\eta -\gamma }\frac{\sqrt{\alpha }A_{r}e^{2\gamma }}{R}\frac{%
V^{2}V^{3}}{V^{0}}= \\
\end{align*}
\begin{align*}
& =2U^{1}\frac{e^{\eta -\gamma }}{V^{0}}\left[ -\left( \sqrt{\alpha }\left(
\eta _{r}-\gamma _{r}\right) +\frac{\alpha _{r}}{2\sqrt{\alpha }}\right)
\left( 1+(V^{2})^{2}+(V^{3})^{2}\right) \right] + \\
& +2U^{1}\frac{e^{\eta -\gamma }}{V^{0}}\left[ \sqrt{\alpha }\gamma
_{r}\left( V^{2}\right) ^{2}-\sqrt{\alpha }\left( \gamma _{r}-\frac{1}{R}%
\right) \left( V^{3}\right) ^{2}\right] + \\
& +2U^{1}\frac{e^{\eta -\gamma }}{V^{0}}\left( \frac{\sqrt{\alpha }%
A_{r}e^{2\gamma }}{R}V^{2}V^{3}-\frac{\alpha _{r}}{2\sqrt{\alpha }}%
(V^{1})^{2}\right) \\
& =T_{1}+T_{2}+T_{3},
\end{align*}

where%
\begin{align*}
T_{1} & =2U^{1}\frac{e^{\eta-\gamma}}{V^{0}}\left[ -\left( \sqrt{\alpha }%
\left( \eta_{r}-\gamma_{r}\right) +\frac{\alpha_{r}}{2\sqrt{\alpha}}\right)
\left( 1+(V^{2})^{2}+(V^{3})^{2}\right) \right] , \\
T_{2} & =2U^{1}\frac{e^{\eta-\gamma}}{V^{0}}\left[ \sqrt{\alpha}\gamma
_{r}\left( V^{2}\right) ^{2}-\sqrt{\alpha}\left( \gamma_{r}-\frac{1}{R}%
\right) \left( V^{3}\right) ^{2}+\frac{\sqrt{\alpha}A_{r}e^{2\gamma}}{R}%
V^{2}V^{3}\right] , \\
T_{3} & =-U^{1}\frac{e^{\eta-\gamma}}{V^{0}}\frac{\alpha_{r}}{\sqrt{\alpha}}%
(V^{1})^{2}.
\end{align*}

Use the constraint equations $\left( \ref{c2}\right) $ and $\left( \ref{c3}%
\right) $ then $T_{1}$ becomes
\begin{align*}
\left| T_{1}\right| & =\left| 2U^{1}\frac{e^{\eta -\gamma }}{V^{0}}\left[
-\left( \sqrt{\alpha }\left( \eta _{r}-\gamma _{r}\right) +\frac{\alpha _{r}%
}{2\sqrt{\alpha }}\right) \left( 1+(V^{2})^{2}+(V^{3})^{2}\right) \right]
\right| = \\
& =\bigg | 2U^{1}\frac{e^{\eta -\gamma }}{V^{0}}\left[ -\sqrt{\alpha }%
\left( R\gamma _{r}^{2}+\frac{R\gamma _{t}^{2}}{\alpha }+\frac{%
A_{r}^{2}e^{4\gamma }}{4R}+\frac{A_{t}^{2}e^{4\gamma }}{4R\alpha }\right) %
\right] \left( 1+(V^{2})^{2}+(V^{3})^{2}\right) + \\
& +2U^{1}\frac{e^{\eta -\gamma }}{V^{0}}\left[ -\sqrt{\alpha }Re^{2\left(
\eta -\gamma \right) }\rho +\sqrt{\alpha }\gamma _{r}-R\sqrt{\alpha }%
e^{2\left( \eta -\gamma \right) }\left( P_{1}-\rho \right) \right] \left(
1+(V^{2})^{2}+(V^{3})^{2}\right)\bigg | = 
\end{align*}
\begin{align*}
& =\bigg | 2U^{1}\frac{e^{\eta -\gamma }}{V^{0}}\left[ -\sqrt{\alpha }%
\left( R\gamma _{r}^{2}+\frac{R\gamma _{t}^{2}}{\alpha }+\frac{%
A_{r}^{2}e^{4\gamma }}{4R}+\frac{A_{t}^{2}e^{4\gamma }}{4R\alpha }-\gamma
_{r}\right) \left( 1+(V^{2})^{2}+(V^{3})^{2}\right) \right] + \\
& -2U^{1}\frac{e^{\eta -\gamma }}{V^{0}}R\sqrt{\alpha }e^{2\left( \eta
-\gamma \right) }P_{1}\left( 1+(V^{2})^{2}+(V^{3})^{2}\right) \bigg | .
\end{align*}

Notice that from step 1 we have bounds on $\left |\eta\right | ,$ $\left |\gamma\right | ,$ $\alpha $ and $\left |A\right |$ and recall that $\left| V^{2}(t)\right| $ and $\left| V^{3}(t)\right| $ are controlled and $\epsilon\leq R(t)\leq C$. The matter terms can be bounded in terms of $Q^{1}(t)$ because
\begin{equation}
P_{1}(t,r)\leq \rho (t,r)=\int_{\mathbb{R}^{3}}f(t,r,v)v^{0}dv\leq C \left\|
f_{0}\right\| \int_{\left | v^{1}\right | \leq Q^{1}}\left |v^{1}\right |dv^{1}\leq C\left(
Q^{1}(t)\right) ^{2}.\label{matter}
\end{equation}
Furthermore notice that 
\begin{equation}
\frac{U^{1}}{V^{0}}\leq C.
\end{equation}
Hence 
\begin{equation*}
\left| T_{1}\right| \leq C_{1}\underset{r\in \lbrack \epsilon 
,\infty )}{\sup }G(s,\text{%
$\cdot$%
})+C_{2}\left( Q^{1}(s)\right) ^{2}.
\end{equation*}%
For $T_{2}$ use $\left( \ref{ax}\right) $ to obtain%
\begin{align*}
\left| T_{2}\right| & =\left| 2U^{1}\frac{e^{\eta -\gamma }}{V^{0}}\left[ 
\sqrt{\alpha }\gamma _{r}\left( V^{2}\right) ^{2}-\sqrt{\alpha }\left(
\gamma _{r}-\frac{1}{R}\right) \left( V^{3}\right) ^{2}+\frac{\sqrt{\alpha }%
A_{r}e^{2\gamma }}{R}V^{2}V^{3}\right] \right| \leq \\
& \leq C\underset{r\in \lbrack \epsilon ,\infty )}{\sup }G(s,\text{%
$\cdot$%
}).
\end{align*}

By using the constraint equation $\left( \ref{c3}\right) $ and equation $%
\left( \ref{ma1}\right) $

\begin{align*}
\left| T_{3}\right| & =\left| 2U^{1}\frac{e^{\eta -\gamma }}{V^{0}}\frac{%
\alpha _{r}}{2\sqrt{\alpha }}(V^{1})^{2}\right| =\left| 2U^{1}\frac{e^{\eta
-\gamma }}{V^{0}}R\sqrt{\alpha }e^{2\left( \eta -\gamma \right) }\left(
P_{1}-\rho \right) (V^{1})^{2}\right| \leq \\
& \leq {\mid }2U^{1}e^{\eta -\gamma }R\sqrt{\alpha }e^{2\left( \eta
-\gamma \right) }\log Q^{1}(s){ \mid }\left| V^{1}\right| \leq C\left(
Q^{1}(s)\right) ^{2}\log Q^{1}(s).
\end{align*}

Add the estimates for $\left| T_{1}\right| ,$ $\left| T_{2}\right| ,$ and $%
\left| T_{3}\right| $ to obtain 
\begin{equation*}
\frac{d}{ds}(U^{1}(s))^{2}\leq C_{1}\underset{r\in \lbrack \epsilon ,\infty )}{\sup }G(s,\text{%
$\cdot$%
})+C_{2}\left( Q^{1}(s)\right) ^{2}\log Q^{1}(s).
\end{equation*}

Integrate the inequality above, note that $V^{1}(t)=e^{-\left( \eta -\gamma
\right) }U^{1}(t)$ and that $\left |\eta\right | $ and $\left |\gamma\right | $, are bounded. Then take the supremum of $\left |V^{1}(t)\right |$ to obtain 
\begin{equation*}
\left( Q^{1}(t)\right) ^{2}\leq C_{1}+C_{2}\int_{t_{0}}^{t}\left[ \left(
Q^{1}(s)\right) ^{2}\log Q^{1}(s)+\underset{r\in \lbrack 
\epsilon,\infty )}{\sup }G(s,\text{%
$\cdot$%
})\right] ds,
\end{equation*}%
and the lemma is proved.
\end{proof}

\begin{remark}
It is necessary to rescale the momentum $V^{1}$ to obtain cancellation of derivatives of the metric in the characteristic equation for $V^{1}$. In \cite{And1} this was not necessary. The important cancellation in \cite{And1} was instead obtained by a careful analysis of the characteristic equation by observing the fact that the terms had the ``right'' sign. However due to the fact that we have control over the metric functions $\gamma$ and $\eta$ this rescaled momentum is essentially the same as $V^{1}$.
\end{remark}
By the estimate $\left( \ref{quad1}\right) $ for $\underset{r\in \lbrack 
\epsilon ,\infty )}{\sup }G(s,$%
$\cdot$%
$)$ and Lemma 3

\begin{equation*}
\Gamma (t)\leq C_{1}+C_{2}\int_{t_{0}}^{t}\Gamma (s)\log \Gamma (s)ds,t\in[t_{0},T),
\end{equation*}%
which is a Gr\"{o}nvall type of inequality. Because of the local existence
theorem we have control over all quantities in the interval $[0,t_{0}].$ Thus%
$\underset{r\in \lbrack \epsilon ,\infty )}{\sup }G(s,$%
$\cdot$%
$)$ and $Q^{1}(t)$ are uniformly bounded on $[0,T)$. Hence it follows that
the matter terms are also uniformly bounded on $[0,T)$ in view of equation $\left (\ref{matter}\right )$ above. Note that due to
assumption $\left( \ref{as1}\right) $ we have control over $\underset{r\in
\lbrack 0,\epsilon )}{\sup }G(s,$%
$\cdot$%
$)$,

and 
\begin{equation*}
\frac{1}{2}(\frac{\gamma _{t}^{2}}{\alpha }+\gamma _{r}^{2})+\frac{%
e^{4\gamma }}{8r^{2}}(\frac{A_{t}^{2}}{\alpha }+A_{r}^{2})=G\leq \underset{%
r\in \lbrack 0,\infty )}{\sup }G(t,\text{%
$\cdot$%
}),
\end{equation*}%
where every term on the left-hand side is nonnegative. This estimate and
that $\alpha \leq \hat{C}$ by $\left( \ref{alf}\right) $,  implies that $%
\left |\gamma _{t}\right |$ and $\left |\gamma _{r}\right |$ are bounded when $r\in[0,\infty)$ and that $\left |A_{t}\right |$ and $\left |A_{r}\right |$ are
bounded on compact intervals $[0,\check{R}]$. Recall that by the asymptotic
flatness conditions, $\left |A_{t}\right |$ and $\left |A_{r}\right |$ vanishes when $r\geq \widetilde{R}+%
\hat{C}t,$ with $\widetilde{R}$ and $\hat{C}$ as before, see the discussion after equation $\left (\ref{alf}\right )$, so that $\left |A_{t}\right |$ and 
$\left |A_{r}\right |$ are bounded everywhere.

The bound on $\left |\alpha _{r}\right |$ follows from the constraint equation $\left( \ref%
{c3}\right) $ and that $\alpha _{r}\equiv 0$ when $r\geq \widetilde{R}+%
\hat{C}t,$ by the asymptotic flatness condition. Hence 
\begin{equation*}
\left| \alpha _{r}\right| =\left| 2r\alpha e^{2\left( \eta -\gamma \right)
}\left( P_{1}-\rho \right) \right| \leq C\log Q^{1}(t).
\end{equation*}

From the constraint equations $\left( \ref{c1}\right) $ and $\left( \ref{c2}%
\right) $ the estimates on $\left |\eta _{t}\right |$ and $\eta _{r}$ follows.
Furthermore $\left |\alpha_{t}\right |\leq C\log Q^{1}(t)\leq C$, see Lemma 6 in the appendix.

\underline{$Step\text{ } 3:$} (Bounds on the first order derivatives of the matter terms.)

To obtain bounds on the first order derivatives of $f,$ it is sufficient to
obtain bounds on $\partial R$ and $\partial V$ because the solution $f$ can
be written in the form, %
\begin{equation*}
f(t,r,v)=f_{0}(R(0,t,r,v),V(0,t,r,v)),
\end{equation*}%
where $\left( R(s,t,r,v),V(s,t,r,v)\right) $ is the solution to the
characteristic system associated to the Vlasov equation, see $\left (\ref{v}\right )$, with $R(t,t,r,v)=r$ and $V(t,t,r,v)=v.$ Note that a partial derivative for $f$ reads

\begin{equation*}
\frac{\partial f}{\partial z}=\frac{\partial f_{0}}{\partial r}\frac{\partial R}{\partial z}+\frac{\partial f_{0}}{\partial v}\frac{\partial V}{\partial z},
\end{equation*}
where $z$ denotes either $t$, $r$ or $v$. This part of the proof is almost identical to the proof of Lemma 3 in \cite{And1} but we have to use assumption $\left (\ref{as2}\right )$ that the matter never reaches the axis in finite time.

\begin{lemma}
Let $R(s)=R(s,t,r,v)$ and $V^{k}(s)=V^{k}(s,t,r,v)$, $k=1,2,3$ be a solution
to the characteristic system associated to the Vlasov equation. Let $%
\partial $ denote $\partial _{t},\partial _{r}$ or $\partial _{v}$ and define%
\begin{align*}
\Psi & =\partial R, \\
Z^{1}& =\partial V^{1}+\left( \frac{\eta _{t}V^{0}}{\sqrt{\alpha }}-\frac{%
\gamma _{t}V^{0}}{\sqrt{\alpha }}\frac{\left( V^{0}\right) ^{2}-\left(
V^{1}\right) ^{2}+\left( V^{2}\right) ^{2}-\left( V^{3}\right) ^{2}}{\left(
V^{0}\right) ^{2}-\left( V^{1}\right) ^{2}}\right) \partial R+ \\
& +\left( \frac{A_{r}e^{2\gamma }}{R}\frac{V^{1}V^{2}V^{3}}{\left(
V^{0}\right) ^{2}-\left( V^{1}\right) ^{2}}-\frac{A_{t}e^{2\gamma }}{\sqrt{%
\alpha }R}\frac{V^{0}V^{2}V^{3}}{\left( V^{0}\right) ^{2}-\left(
V^{1}\right) ^{2}}\right) \partial R+ \\
& \gamma _{r}\frac{V^{1}\left( \left( V^{2}\right) ^{2}-\left( V^{3}\right)
^{2}\right) }{\left( V^{0}\right) ^{2}-\left( V^{1}\right) ^{2}}\partial R,
\\
\end{align*}
\begin{align*}
Z^{2}& =\partial V^{2}+\gamma _{r}V^{2}\partial R, \\
Z^{3}& =\partial V^{3}+\left( \frac{A_{r}e^{2\gamma }V^{2}}{R}-\gamma
_{r}V^{3}\right) \partial R.
\end{align*}

Then there is a matrix $A=\left( a_{lm}\right) ,l,m=0,1,2,3$ such that%
\begin{equation*}
\Omega\equiv(\Psi,Z^{1},Z^{2},Z^{3})^{T},
\end{equation*}

satisfies%
\begin{equation}
\frac{d\Omega }{ds}=A\Omega ,  \label{om}
\end{equation}

and the matrix elements are all uniformly bounded on $[0,T).$
\end{lemma}

The proof is obtained by a long calculation, see \cite{And1} for a sketch of the proof.

By standard results from ordinary differential equations $\Omega $ is uniformly
bounded on $[0,T)$, and hence the derivatives of the matter terms are
uniformly bounded on $[0,T).$ It is then immediate to obtain bounds on $\partial R$ and $\partial V^{i}, i=1,2,3$.

\underline{$Step\text{ } 4:$} (Bounds on second order and higher order derivatives.)

This section also follows \cite{And1} and the methods used in step 2. Hence derivatives of arbitrary order both for the field and the matter can be uniformly bounded on $[0,T)$, so the solution can be extended beyond $T$ which is a contradiction. Hence $T=\infty $ and the proof is complete.

\section{Proof of theorem 1}
\begin{proof}
Given $\epsilon>0$. If $T=\infty$ we are done so assume that $T<\infty$. If we are able to obtain uniform bounds on all the functions (including their derivatives) we can extend the solution $(f,g_{ab})$ to $t=T$, and then apply the local existence theorem again with initial data at $t=T$, contradicting the fact that $[0,T)$ was the maximal time of existence. To exploit assumption $\left(\ref{a}\right )$ we divide $([0,T)\times\mathbb{R}^{3},g_{ab})$ into an interior-, $r<\frac{\epsilon}{4}$, and an exterior region, where $r\geq\frac{\epsilon}{4}$. The bounds in the exterior region rely on the proof of Theorem 2. However in that proof the assumption $\left (\ref{as1}\right )$ that the derivatives of $\gamma$ are controlled in the interior region was used in a crucial way. By an examination of the light-cone argument in the proof of Theorem 2 we see that it is necessary to obtain an estimate of the type $\left |\partial\gamma\right |\leq Q^{1}(t)$ in the interior region, and then apply the light cone argument as in the proof of Theorem 2 to the exterior region. To obtain this ``interior'' estimate we use a standard Sobolev method.

From the proof of Theorem 2 we have that $\alpha$ and $|\eta |$ are uniformly bounded on $[0,T)\times[0,\infty)$. Furthermore $\left |\gamma\right |$ is uniformly bounded on $[0,T)\times[\epsilon,\infty)$, for any $\epsilon >0$. From $\left (\ref{K1}\right )$ and $\left (\ref{K2}\right )$ we have

\begin{align*}
V^{2}(t)e^{\gamma(t,R(t))}=C,\\
V^{3}(t)R(t)e^{-\gamma(t,R(t))}=C.
\end{align*}
The discussion before and after equation $\left (\ref{ax}\right )$ yields that
\begin{align*}
\left | V^{2}(t)\right |\leq C,\\
\left | V^{3}(t)\right |\leq C.
\end{align*}
Hence
\begin{equation}
\sup\{\left |v^{2}\right |+\left |v^{3}\right | +3:\exists (s,r,v^{1})\in [0,t)\times[0,\infty)\times(-\infty,\infty) \text{ with } f(s,r,v)\neq 0\},
\end{equation}
is also uniformly bounded on $[0,T)$. So in order to control $Q(t)$, see equation $\left (\ref{Q}\right )$, it is enough to control $Q^{1}(t):=\sup\{\left |v^{1}\right |+3:\exists (s,r,v^{2},v^{3})\in[0,t)\times[0,\infty)\times\mathbb{R}^{2} \text{ with } f(s,r,v)\neq 0\}$.
By the constraint equation $\left (\ref{c3}\right )$ and assumption $\left (\ref{a}\right )$
\begin{equation}
\alpha_{r}(t,r)\equiv 0, t\in [0,T), r<\epsilon.
\end{equation}
Hence $\alpha=\alpha(t,r)$ is independent of $r$ when $r<\epsilon$. Observe from the discussion above that assumption $\left (\ref{a}\right )$ implies that the matter terms vanish when $r<\epsilon$.

Now $\gamma$ is a rotationally symmetric solution of
\begin{equation}
\Box_{H}\gamma=F, \label{w}
\end{equation}
where
\begin{equation}
\Box_{H}=-\frac{\partial_{t}^{2}}{\alpha}+\partial_{x}^{2}+\partial_{y}^{2}+\frac{\alpha_{t}}{2\alpha^{2}}\partial_{t}+\frac{\alpha_{x}}{2\alpha}\partial_{x}+\frac{\alpha_{y}}{2\alpha}\partial_{y},
\end{equation}
with $x=r\cos\theta$, $y=r\sin\theta$, and
\begin{equation}
F=-\frac{e^{2(\eta-\gamma)}}{2}(\rho-P_{1}+P_{2}-P_{3}).
\end{equation}

Let
\begin{equation}
\delta:=\frac{\epsilon}{4}.
\end{equation}
Define the backwards truncated ``cone'' $\Omega =\{(t,x,y):t_{0}\leq t\leq T,\sqrt{x^{2}+y^{2}}\leq \check{r}(t)\}$, where $\check{r}(s)$ is the unique solution to 
\begin{equation}
\frac{d\check{r}}{ds}=-\sqrt{\alpha(s,\check{r}(s))},\text{ }\check{r}(T)=\delta.
\end{equation}

Define

\begin{equation}
B_{t}:=\{(x,y):\sqrt{x^{2}+y^{2}}\leq T+\delta-t\},
\end{equation}

and let $C_{t_{0},t}$ be the surface of the cone.
 Choose $t_{0}$ big enough such that the radius of $B_{t_{0}}$ is less than $2\delta$. This is always possible, and in view of the local existence theorem due to Choquet-Bruhat \cite{CB} we can control all quantities on the hypersurface $t=t_{0}$.
Then we can use the following special case of lemma 2 from \cite{K}.

\begin{lemma} Let $u\in H_{loc}^{2}(\mathbb{R}^{n})$ with $\tilde{B}_{\delta}(\vec{x})=\{\vec{y}\in\mathbb{R}^{n}:\left |\vec{y}-\vec{x}\right |<\delta\}$ an arbitrary ball. Then with $|\vec{x}|$,
\begin{equation}
\left |u(\vec{x})\right |\leq c\sum_{k=0}^{2}\delta^{k-1}\parallel\partial^{k}u\parallel_{L^{2}(\tilde{B}_{\delta}(\vec{0}))}.
\end{equation}
\end{lemma}

\begin{remark}
Define $\mu_{H}:=\sqrt{\alpha}dxdy$ as the measure associated to the metric $H_{ij}$. That $\sqrt{\alpha}\geq 1$ in our case is immediate from the constraint equation $\left( \ref{c3}\right) $, see $\left(\ref{alf}\right)$. 
\end{remark}

Hence
\begin{equation}
|\partial\gamma(t,x,y) |\leq C\sum_{\nu=0}^{2}\parallel\partial^{\nu}\partial\gamma(t,\cdot,\cdot)\parallel_{L^{2}({B_{t},\mu_{H}})}.
\end{equation}

A typical term to be estimated in the Sobolev inequality above reads
\begin{equation}
\parallel\partial^{\nu}\partial\gamma(t,\cdot,\cdot)\parallel_{L^{2}(B_{t},\mu_{H})},
\end{equation}
for $0\leq\nu\leq 2$. Observe that $\partial^{\nu}=\partial^{\nu_{1}}_{x}\partial^{\nu_{2}}_{y}$ where $\nu_{1}+\nu_{2}=\nu$. Note also that $\frac{1}{2}(\gamma_{t}^{2}+\alpha\gamma_{x}^{2}+\alpha\gamma_{y}^{2})-\sqrt{\alpha}\frac{x}{r}\gamma_{t}\gamma_{x}-\sqrt{\alpha}\frac{y}{r}\gamma_{t}\gamma_{y}\geq 0$ and apply the local energy identity which reads

\begin{equation}
\parallel\partial^{\nu}\partial\gamma\parallel_{L^{2}(B_{t},\mu_{H})}^{2}=
\end{equation}
\begin{equation}
=\parallel\partial^{\nu}\partial\gamma\parallel_{L^{2}(B_{t_{0}},\mu_{H})}^{2}-\int_{C_{t_{0},t}}[\frac{1}{2}(\gamma_{t}^{2}+\alpha\gamma_{x}^{2}+\alpha\gamma_{y}^{2})-\sqrt{\alpha}\frac{x}{r}\gamma_{t}\gamma_{x}-\sqrt{\alpha}\frac{y}{r}\gamma_{t}\gamma_{y}]d\sigma+
\end{equation}
\begin{equation}
-\frac{1}{2}\int_{\Omega}\frac{\alpha_{t}}{2\alpha^{2}}(\gamma_{t}^{2}+\alpha\gamma_{x}^{2}+\alpha\gamma_{y}^{2})\sqrt{\alpha}dxdyd\tau-\int_{\Omega}\partial^{\nu}F\partial^{\nu}\gamma_{t}\sqrt{\alpha}dxdyd\tau\leq
\end{equation}
\begin{equation}
\leq\parallel\partial^{\nu}\partial\gamma\parallel_{L^{2}(B_{t_{0}},\mu_{H})}^{2}+C\underset{(x,y)\in\mathbb{R}^{2}}{\sup}\left |\alpha_{t}(t,\cdot,\cdot)\right |\int_{t_{0}}^{t}\parallel\partial^{\nu}\partial\gamma\parallel_{L^{2}(B_{\tau},\mu_{H})}^{2}d\tau,
\end{equation}

where $d\sigma$ is the induced measure on the surface of the cone.

So we obtain a Grönvall inequality for $\parallel\partial^{\nu}\partial\gamma\parallel_{L^{2}(B_{t},\mu_{H})}^{2}$. Hence
\begin{equation}
\parallel\partial^{\nu}\partial\gamma(t,\cdot,\cdot)\parallel^{2}_{L^{2}(B_{t},\mu_{H})}\leq C_{1}e^{(t-t_{0})\underset{\tau\in[t_{0},t], (x,y)\in\mathbb{R}^{2}}\sup\left |\alpha_{t}(\cdot,\cdot,\cdot)\right |}.
\end{equation}
From Lemma 6 in the appendix we get that $\left |\alpha_{t}\right |$ can be estimated by
\begin{equation}
\left |\alpha_{t}(t,x,y)\right |\leq C_{2}\log{Q^{1}(t)}.
\end{equation}

Finally
\begin{equation}
\parallel\partial^{\nu}\partial\gamma(t,\cdot,\cdot)\parallel^{2}_{L^{2}(B_{t},\mu_{H})}\leq C_{1}\left [Q^{1}(t)\right ]^{C_{2}(t-t_{0})}.
\end{equation}
From Lemma 6 in appendix we see that $C_{2}$ do not depend on $t_{0}$, so by choosing $t_{0}$ such that $C_{2}(t-t_{0})<1\label{t0}$ and of course such that the condition that the radius of $B_{t_{0}}$ is less than $2\delta$ holds, then

\begin{equation}
\left |\partial\gamma(t,x,y)\right |^{2}\leq CQ^{1}(t), t\in[t_{0},T), \sqrt{x^{2}+y^{2}}\leq\delta.\label{GN}
\end{equation}

Now we can rely on the proof of Theorem 2 with $A\equiv 0$ and with $\epsilon$ replaced by $\frac{\epsilon}{4}$ in that proof. Furthermore observe that we have to replace assumption $\left (\ref{as1}\right )$ by the estimate
\begin{equation}
\underset{r\in[\frac{\epsilon}{8},\frac{\epsilon}{4}]}{\sup }G(t,\cdot)\leq CQ^{1}(t),
\end{equation}
which was obtained above. Recall the definition of $G(t,r)$ and $H(t,r)$
\begin{equation}
G:=\frac{1}{2}\left (\frac{\gamma_{t}^{2}}{\alpha}+\gamma_{r}^{2}\right ),
\end{equation}
\begin{equation}
H:=\frac{\gamma_{t}\gamma_{r}}{\sqrt{\alpha}}.
\end{equation}
By using $\left (\ref{GN}\right )$ instead of assumption $\left (\ref{as1}\right )$, after choosing $t_{0}$ as in the discussion after equation $\left (\ref{t}\right )$ and of course such that the radius of $B_{t_{0}}$ is less than $2\delta$ holds and also that $\left (\ref{GN}\right )$ hold. Then the light cone estimate in the proof of Theorem 2 becomes, see equation $\left (\ref{quad1}\right )$,

\begin{equation*}
\underset{r\in\left [\frac{\epsilon}{4},\infty\right  )}{\sup }G\left ( t,\cdot\right )\leq
\end{equation*}

\begin{equation*}
\underset{r\in \left [ \frac{\epsilon}{8},\infty\right )}{\sup }G\left (t_{0},\cdot\right )+\underset{r\in \left [ \frac{\epsilon}{8},\infty \right )}{\sup }H\left (t_{0},\cdot\right )+
\end{equation*} 

\begin{equation*}
+\underset{r\in \left [ \frac{\epsilon }{8},\infty\right )}{\sup }\int_{t_{0}}^{t}C\frac{\kappa e^{2\left( \eta -\gamma \right ) }}{2}\sqrt{G(s,\cdot)}ds+\underset{r\in \left [ \frac{\epsilon }{8},\infty\right )}{\sup }\int_{t_{0}}^{t}Cr\sqrt{\alpha}e^{2(\eta-\gamma)}\log Q^{1}(s)\left ( G(s,\cdot)-H(s,\cdot)\right )ds+
\end{equation*}

\begin{equation*}
+\underset{r\in \left [ \frac{\epsilon }{8},\infty\right )}{\sup }\int_{t_{0}}^{t}\sqrt{\alpha} \frac{C}{r}G(s,\cdot)ds
\end{equation*}

\begin{align*}
\leq C_{1}+C_{2}(\epsilon)\int_{t_{0}}^{t}\log Q^{1}(s)\underset{r\in \lbrack 
\frac{\epsilon}{4} ,\infty )}{\sup }G(s,\text{%
$\cdot$%
})ds+C_{3}(\epsilon)\int_{t_{0}}^{t}\underset{r\in \lbrack 
\frac{\epsilon}{4} ,\infty )}{\sup }G(s,\text{%
$\cdot$%
})ds+
\end{align*}

\begin{align*}
C_{4}(\epsilon)\int_{t_{0}}^{t}\log Q^{1}(s)\underset{r\in \lbrack 
\frac{\epsilon}{8} ,\frac{\epsilon}{4} )}{\sup }G(s,\text{%
$\cdot$%
})ds+C_{5}(\epsilon)\int_{t_{0}}^{t}\underset{r\in\left [\frac{\epsilon}{8},\frac{\epsilon}{4}\right )}{\sup}G(s,\cdot)ds\leq
\end{align*}

\begin{align*}
\leq C_{1}+C_{2}(\epsilon)\int_{t_{0}}^{t}\log Q^{1}(s)\underset{r\in \lbrack 
\frac{\epsilon}{4} ,\infty )}{\sup }G(s,\text{%
$\cdot$%
})ds+C_{3}(\epsilon )\int_{t_{0}}^{t}\underset{r\in \lbrack 
\frac{\epsilon}{4} ,\infty )}{\sup }G(s,\text{%
$\cdot$%
})ds+
\end{align*}%

\begin{align*}
+C_{4}(\epsilon)\int_{t_{0}}^{t}Q^{1}(s)\log Q^{1}(s)ds+C_{5}(\epsilon)\int_{t_{0}}^{t}Q^{1}(s)ds.
\end{align*}

By combining this with Lemma 3, which holds without any changes, a Grönvall type inequality is obtained for
\begin{equation}
\Gamma(s):=\underset{r\in [\frac{\epsilon}{4},\infty)}{\sup}G(t,\cdot)+\left (Q^{1}(t)\right )^{2}.
\end{equation}
It reads
\begin{equation}
\Gamma(t)\leq C_{1}+C_{2}\int_{t_{0}}^{t}\Gamma (s)\log\Gamma (s)ds.
\end{equation}
Hence $\Gamma(t)$ is uniformly bounded on $[t_{0},T)$ and as mentioned above we have control over all quantities when $t\in[0,t_{0}]$. So $\Gamma(t)$ is uniformly bounded on $[0,T)$. Together with equation $\left (\ref{GN}\right )$ this implies that $G$ is uniformly bounded on $[0,T)\times [0,\infty)$.

To conclude we obtain uniform bounds on the metric functions and all the first order derivatives and on the matter terms, see the proof of Theorem 2 and the appendix.
Now we can continue in the same way as in the proof of Theorem 2 with the only modification that $\epsilon$ is replaced by $\frac{\epsilon}{4}$ and that assumption $\left (\ref{as1}\right )$ is replaced by
\begin{equation}
\left |\partial^{\beta}{\gamma}\right |^{2}\leq CQ^{1}(t),\forall\beta\geq 1, t\in[t_{0},T), \sqrt{x^{2}+y^{2}}<\delta,
\end{equation}
where $Q^{1}(t)\leq C$ in view of the estimates above because $\alpha$ is independent of $r$ when $r<\epsilon$.
Hence Theorem 2 is proved.

\end{proof}
 \begin{remark}
In the nonpolarized case this Sobolev method does not work. This is to be expected in view of the deep proof of Christodoulou and Tahvildar-Zadeh on the spherically symmetric wave maps from flat space into a class of Riemannian manifolds. 
\end{remark}

 \section{Appendix}
\begin{lemma} With all the functions as above the following estimate holds
\begin{equation}
|\alpha_{t}(t,r)|\leq C\log Q^{1}(t), t\in [0,T),
\end{equation}
where the constant $C$ only depend on the maximal time of existence $T$.
\end{lemma}
\begin{proof}
Observe that we have uniform bounds on $\alpha$ and $\left |\eta\right |$ when $\left (t,r\right )\in[0,T)\times[0,\infty)$ and a uniform bound on $\left |\gamma\right |$ when $\left (t,r\right )\in[0,T)\times[\epsilon,\infty)$, see step 1 in the proof of Theorem 2. Furthermore assumption $\left (\ref{a}\right )$ implies that $\left |V^{2}(t)\right |$ and $\left |V^{3}(t)\right |$ are uniformly bounded in time, see step 2 in the proof of Theorem 2.
Now integrate the constraint equation $\left (\ref{c3}\right )$
\begin{equation*}
\frac{\alpha_{r}(t,r)}{\alpha(t,r)}=2re^{2(\eta-\gamma)}(P_{1}-\rho).
\end{equation*}
Use the asymptotic flatness condition $\left (\ref{b4}\right )$
\begin{equation*}
\underset{r\rightarrow\infty}{\lim}\alpha(t,r)=1,
\end{equation*}
to get
\begin{equation*}
\log{\alpha(t,r)}=2\int_{r}^{\infty}r^{\prime}e^{2(\eta-\gamma)}(\rho-P_{1})dr^{\prime}.
\end{equation*}
Hence
\begin{equation*}
\frac{\alpha_{t}(t,r)}{\alpha(t,r)}=2\int_{r}^{\infty}r^{\prime}e^{2(\eta-\gamma)}\left [2(\eta_{t}-\gamma_{t})(\rho-P_{1})+\partial_{t}(\rho-P_{1})\right ]dr^{\prime}=
\end{equation*}
\begin{equation*}
=2\int_{0}^{\infty}re^{2(\eta-\gamma)}\left [2(\eta_{t}-\gamma_{t})(\rho-P_{1})+\partial_{t}(\rho-P_{1})\right ]dr.
\end{equation*}

Observe that assumption $\left (\ref{a}\right )$ implies that $\alpha_{t}=\alpha_{t}(t),\text{ }r<\epsilon$. 
The first term in the integral above can be estimated as follows
\begin{equation*}
\bigg | 4\int_{0}^{\infty}re^{2(\eta-\gamma)}\left (\eta_{t}-\gamma_{t}\right )(\rho-P_{1})dr\bigg | \leq C\log{Q^{1}(t)}\int_{0}^{\infty}r\left |\eta_{t}-\gamma_{t}\right |dr\leq
\end{equation*}
\begin{equation*}
\leq C\log{Q^{1}(t)}\int_{0}^{\infty}r\left (\left |\eta_{t}\right |+\left |\gamma_{t}\right |\right )dr.
\end{equation*}
From the constraint equations $\left (\ref{c1}\right )$ and $\left (\ref{c2}\right )$ for $\eta$
\begin{equation*}
\left |\eta_{t}\right |\leq C\eta_{r}.
\end{equation*}
Recall that the derivatives of the metric have compact support so by Hölder's inequality  
\begin{equation*}
C\log{Q^{1}(t)}\int_{0}^{\infty}r\left (\left |\eta_{t}\right |+\left |\gamma_{t}\right |\right )dr\leq C\log{Q^{1}(t)}\int_{0}^{\infty}r\left (\eta_{r}+\left |\gamma_{t}\right |\right )dr\leq
\end{equation*}
\begin{align*}
\leq C_{1}\log{Q^{1}(t)}\int_{0}^{\infty}\eta_{r}dr+C_{2}\log{Q^{1}(t)}\int_{0}^{\infty}r\left |\gamma_{t}\right |dr\leq \\
\leq C\log{Q^{1}(t)}\left [\int_{0}^{\infty}\eta_{r}dr+\left (\int_{0}^{\infty}\left |\gamma_{t}\right |^{2}rdr\right )^{\frac{1}{2}}\left (\int_{0}^{R+\hat{C}t}rdr\right )^{\frac{1}{2}}\right ]\leq 
\end{align*}
\begin{equation*}
\leq C\log{Q^{1}(t)},
\end{equation*}
 where we used that
\begin{equation*}
\int_{0}^{\infty}\left |\gamma_{t}\right |^{2}rdr\leq\int_{0}^{\infty}\eta_{r}dr\leq C.
\end{equation*}
Hence
\begin{equation*}
\bigg | 4\int_{0}^{\infty}re^{2(\eta-\gamma)}\left (\eta_{t}-\gamma_{t}\right )(\rho-P_{1})dr \bigg | \leq C\log{Q^{1}(t)}
\end{equation*}
For the second term in the integral expression of $\alpha_{t}$ above use the Vlasov equation to obtain
\begin{equation*}
\left |2\int_{0}^{\infty}re^{2(\eta-\gamma)}\partial_{t}(\rho-P_{1})dr\right |=\bigg | 2\int_{0}^{\infty}\int_{\mathbb{R}^{3}}re^{2(\eta-\gamma)}\frac{1+(v^{2})^{2}+(v^{3})^{2}}{v^{0}}\biggl [ -\sqrt{\alpha}\frac{v^{1}}{v^{0}}\frac{\partial f}{\partial r}+
\end{equation*}
\begin{equation*}
+\left (\sqrt{\alpha}(\eta_{r}-\gamma_{r})+\frac{\alpha_{r}}{2\sqrt{\alpha}}\right )v^{0}\frac{\partial f}{\partial v^{1}}+\left (-\sqrt{\alpha}\gamma_{r}\frac{(v^{2})^{2}}{v^{0}}+\sqrt{\alpha}(\gamma_{r}-\frac{1}{r})\frac{(v^{3})^{2}}{v^{0}}+(\eta_{t}-\gamma_{t})v^{1}\right )\frac{\partial f}{\partial v^{1}}+
\end{equation*}
\begin{equation*}
+\left (\gamma_{t}v^{2}+\sqrt{\alpha}\gamma_{r}\frac{v^{1}v^{2}}{v^{0}}\right ) \frac{\partial f}{\partial v^{2}}+\left (\sqrt{\alpha}(\frac{1}{r}-\gamma_{r})\frac{v^{1}v^{3}}{v^{0}}-\gamma_{t}v^{3}\right )\frac{\partial f}{\partial v^{3}} \biggr ] dvdr\bigg | .
\end{equation*}
We estimate different terms separately as follows. By the conserved quantity $\left (\ref{energy}\right )$, the observation in the beginning of appendix and partial integration
\begin{equation*}
\left |2\int_{0}^{\infty}\int_{\mathbb{R}^{3}}re^{2(\eta-\gamma)}\frac{1+(v^{2})^{2}+(v^{3})^{2}}{v^{0}}\sqrt{\alpha}\frac{v^{1}}{v^{0}}\frac{\partial f}{\partial r}dvdr\right |=
\end{equation*}

\begin{equation*}
=\bigg | 2re^{2(\eta-\gamma)}\sqrt{\alpha}\int_{\mathbb{R}^{3}}\frac{1+(v^{2})^{2}+(v^{3})^{2}}{v^{0}}\frac{v^{1}}{v^{0}}f(t,r,v)dv+
\end{equation*}
\begin{equation*}
+2\int_{0}^{\infty}\int_{\mathbb{R}^{3}}\frac{1+(v^{2})^{2}+(v^{3})^{2}}{v^{0}}\frac{v^{1}}{v^{0}}f(t,r,v)\biggl [ e^{2(\eta-\gamma)}\sqrt{\alpha}+
\end{equation*}
\begin{equation*}
+r[e^{2(\eta-\gamma)}\sqrt{\alpha}2(\eta_{r}-\gamma_{r})+re^{2(\eta-\gamma)}\frac{\alpha_{r}}{2\sqrt{\alpha}}]\biggr ] dvdr\bigg | \leq 
\end{equation*}
\begin{equation*}
\leq C_{1}\log{Q^{1}(t)}+2\int_{0}^{\infty}\int_{\mathbb{R}^{3}}\frac{1+(v^{2})^{2}+(v^{3})^{2}}{v^{0}}\frac{v^{1}}{v^{0}}f(t,r,v)r^{\prime}e^{2(\eta-\gamma)}\frac{\alpha_{r}}{2\sqrt{\alpha}}dvdr^{\prime}\leq
\end{equation*}
\begin{equation*}
\leq C_{1}\log{Q^{1}(t)}+C_{2}\log{Q^{1}(t)}\int_{0}^{\infty}\left |\frac{\alpha_{r}}{2\sqrt{\alpha}}\right |dr\leq  C_{1}\log{Q^{1}(t)}+
\end{equation*}
\begin{equation*}
C_{2}\log{Q^{1}(t)}\int_{0}^{\infty} re^{2(\eta-\gamma)}\rho dr\leq 
\end{equation*}

\begin{equation*}
C\log{Q^{1}(t)}.
\end{equation*}

By partial integration
\begin{equation*}
\left |2\int_{0}^{\infty}\int_{\mathbb{R}^{3}}re^{2(\eta-\gamma)}\frac{1+(v^{2})^{2}+(v^{3})^{2}}{v^{0}}\left (\sqrt{\alpha}(\eta_{r}-\gamma_{r})+\frac{\alpha_{r}}{2\sqrt{\alpha}}\right )v^{0}\frac{\partial f}{\partial v^{1}}dvdr\right |=0.
\end{equation*}
Again by partial integration
\begin{equation*}
\bigg | 2\int_{0}^{\infty}\int_{\mathbb{R}^{3}}re^{2(\eta-\gamma)}\frac{1+(v^{2})^{2}+(v^{3})^{2}}{v^{0}}\left (-\sqrt{\alpha}\gamma_{r}\frac{(v^{2})^{2}}{v^{0}}+\sqrt{\alpha}(\gamma_{r}-\frac{1}{r})\frac{(v^{3})^{2}}{v^{0}}\right )\frac{\partial f}{\partial v^{1}}dvdr+
\end{equation*}
\begin{equation*}
+2\int_{0}^{\infty}\int_{\mathbb{R}^{3}}re^{2(\eta-\gamma)}\frac{1+(v^{2})^{2}+(v^{3})^{2}}{v^{0}}\left (\eta_{t}-\gamma_{t}\right )v^{1}\frac{\partial f}{\partial v^{1}}dvdr\bigg | =
\end{equation*}
\begin{equation*}
=\bigg | 2\int_{0}^{\infty}\int_{\mathbb{R}^{3}}re^{2(\eta-\gamma)}f(t,r,v)\left (1+(v^{2})^{2}+(v^{3})^{2}\right )(-\sqrt{\alpha}\gamma_{r}(v^{2})^{2}\frac{2v^{1}}{(v^{0})^{4}}+
\end{equation*}
\begin{equation*}
+\sqrt{\alpha}(\gamma_{r}-\frac{1}{r})(v^{3})^{2}\frac{2v^{1}}{(v^{0})^{4}}+(\eta_{t}-\gamma_{t})\left (-\frac{1}{v^{0}}+\frac{(v^{1})^{2}}{(v^{0})^{3}}\right )dvdr\bigg | \leq C\log{Q^{1}(t)}.
\end{equation*}
The last terms can be estimated in a similar way by partial integration. Sum all estimates to obtain
\begin{equation*}
\left |\alpha_{t}(t,r)\right |\leq C\log{Q^{1}(t)}.
\end{equation*}
Furthermore it is immediate consequence from the proof above that $C$ only depends on the maximal time of existence $T$.
\end{proof}

\textbf{Acknowledgement:} I want to thank Håkan Andreasson for suggesting the problem and for useful discussions and comments.

\end{document}